\documentclass[12pt,a4paper]{article}
\usepackage[utf8]{inputenc}
\usepackage{amsmath,graphicx,dcolumn,subcaption,mathtools,amssymb}
\usepackage{braket}
\usepackage{siunitx}
\usepackage{appendix}
\usepackage[mathscr]{eucal}
\usepackage{bm}
\usepackage{bbm}
\usepackage{cancel}
\usepackage{color}
\usepackage{rotating}
\usepackage{xcolor}
\usepackage[most]{tcolorbox}

\definecolor{darkgreen}{RGB}{0,100,0}

\usepackage[numbers,sort&compress]{natbib}
\usepackage[]{float}
\usepackage[font=small, labelfont=bf]{caption}

\usepackage{slantsc}
\usepackage{xspace}

\usepackage{setspace}
\allowdisplaybreaks

\newcommand\asbare{\alpha_{\mathrm{S}}^u} 
\newcommand\as{\alpha_{\mathrm{S}}} 
 
\newcommand\eps{\epsilon}

\def\beq{\begin{equation}} 
\def\eeq{\end{equation}} 
\def\beeq{\begin{eqnarray}} 
\def\eeeq{\end{eqnarray}}

\newcommand{\mR}{\mathcal{R}}
\newcommand{\mJ}{\mathcal{J}}
\newcommand{\mS}{\mathcal{S}}
\newcommand{\mSsub}{\mathcal{S_{\rm sub}}}
\newcommand{\mJS}{ \mathcal{J} \! \! \!  \mathcal{S}}
\newcommand{\azero}{\alpha _0}

\newcommand{\rcut}{r_{\mathrm{cut}}}

\newcommand\ep{\epsilon}

\usepackage{hyperref}
\usepackage{footnotebackref}

\newcommand{\qfcut}{\mathfrak{q}_{\mathrm{cut}}}
\newcommand{\J}{\mathcal{J}}

\newcommand{\qf}{\mathfrak{q}}

\newcommand{\JJ}{\mathbf{J}\! \!\mathbf{J}}

\newcommand{\twoj}{{e^+e^-\to 2\,{\rm jets}}}
\newcommand{\hbb}{{H\to b{\bar b}}}

\newcommand{\MSbar}{\overline{\mathrm{MS} }}

\begin{document}
\hypersetup{pageanchor=false}
\begin{titlepage}
\begin{flushright}
  ZU-TH 83/25\\
  CERN-TH-2025-232\\
\end{flushright}

\renewcommand{\thefootnote}{\fnsymbol{footnote}}
\vspace*{0.5cm}

\begin{center}
  {\Large \bf Jet Production at NNLO: Exploring a New Scheme}
\end{center}

\par \vspace{2mm}
\begin{center}
  {\bf Luca Buonocore}$^{(a)}$, {\bf Massimiliano Grazzini}$^{(b)}$, {\bf Flavio Guadagni}$^{(b)}$\\[0.2cm]
  {\bf J\"urg Haag}$^{(c)}$, {\bf Stefan Kallweit}$^{(b)}$ and {\bf Luca Rottoli}$^{(d)}$

\vspace{5mm}

$^{(a)}$ CERN, Theoretical Physics Department, CH-1211 Geneva 23, Switzerland\\[0.2cm]
$^{(b)}$ Physik Institut, Universit\"at Z\"urich, CH-8057 Z\"urich, Switzerland\\[0.2cm]
$^{(c)}$ Albert Einstein Center for Fundamental Physics, Institut f\"ur Theoretische Physik, Universit\"at
Bern, Sidlerstrasse 5, CH-3012 Bern, Switzerland\\[0.2cm]
$^{(d)}$ Dipartimento di Fisica G. Occhialini, Universit\`a degli Studi di Milano-Bicocca
and INFN, Sezione di Milano-Bicocca, Piazza della Scienza 3,20126 Milano, Italy\\[0.2cm]
\vspace{5mm}

\end{center}

\par \vspace{2mm}
\begin{center} {\large \bf Abstract} 

\end{center}
\begin{quote}
  \pretolerance 10000

We consider dijet production in $e^+e^-$ collisions and in \mbox{$\hbb$} decays at next-to-next-to-leading order (NNLO) in perturbative QCD. A new non-local subtraction scheme is applied, for the first time, to obtain the fully differential cross section for these benchmark processes. We discuss and explicitly evaluate the perturbative ingredients needed in the computation, and we compare the performance of different slicing variables to obtain the NNLO corrections.

\end{quote}

\vspace*{\fill}
\begin{flushleft}
December 2025
\end{flushleft}
\end{titlepage}

\clearpage
\pagenumbering{arabic}
\setcounter{page}{1}
\hypersetup{pageanchor=true}


\renewcommand{\thefootnote}{\fnsymbol{footnote}}

\section{Introduction}
\label{sec:intro}

Jets play a central role in the study of Quantum Chromodynamics (QCD).
They arise when final-state partons produced in particle collisions fragment into collimated sprays of hadrons and are ubiquitous at high-energy particle colliders.
Jets serve as essential tools in a wide variety of physics analyses (see Ref.~\cite{Stagnitto:2025air} for a recent review).
A prerequisite for the success of these studies is an unambiguous definition of a jet, i.e. a prescription for clustering particles into jets and assigning them well-defined momenta.
To facilitate the comparison with theory, such a definition should be \emph{infrared safe}~\cite{Sterman:1977wj}.
Once an appropriate \emph{jet algorithm} has been introduced, jets can be consistently identified both in parton-level calculations and parton-shower Monte Carlo simulations, as well as in experimental data, thereby providing a well-defined representation of the hadronic final-state in high-energy particle collisions.

In the last 15 years, the Large Hadron Collider (LHC) has been the primary facility to explore the high-energy frontier.
As a consequence, most of the theoretical effort in the last couple of decades has focussed on understanding QCD in hadron--hadron collisions.
The precise description of events with one or more jets in the final state, which constitute the vast majority of those recorded at the LHC, poses conceptual and technical challenges.
Indeed, having three or more coloured legs at tree level significantly increases the complexity of higher-order QCD calculations due to the underlying singularity structure.
A next-to-next-to-leading order (NNLO) description of dijet production, the simplest jet production process at the LHC, has been obtained only recently by two groups~\cite{Currie:2016bfm,Currie:2017eqf,Gehrmann-DeRidder:2019ibf,Czakon:2019tmo,Chen:2022tpk}, whereas only one calculation for trijet production at NNLO exists to date~\cite{Czakon:2021mjy}.

All these results are based on \emph{local} subtraction schemes~\cite{Gehrmann-DeRidder:2005btv,Daleo:2006xa,Currie:2013vh,Czakon:2010td,Czakon:2011ve,Czakon:2014oma}, which rely on suitable counterterms to make all contributions finite and integrable in the singular regions.
An alternative approach to achieving NNLO accuracy is provided by \emph{non-local} or \emph{slicing} schemes, which regulate divergent integrals by introducing a variable sensitive to infrared radiation and splitting the phase space into unresolved and (partially) resolved regions.
These methods base the computation of NNLO corrections on a next-to-leading order (NLO) calculation with higher multiplicity, thereby exploiting the extensive experience developed for NLO computations.
For these schemes to succeed, it is essential to know the singular structure of the higher-multiplicity NLO cross section when an appropriate resolution variable vanishes.
A well-known candidate for such a variable is $N$-jettiness $\tau_N$~\cite{Stewart:2010tn,Gaunt:2015pea}.
NNLO predictions with $N$-jettiness slicing have been obtained for the associated production of a boson and a jet in Refs.~\cite{Boughezal:2015aha,Boughezal:2015ded,Boughezal:2016dtm,Alioli:2025hpa}, and the additional theoretical ingredients required to reach NNLO accuracy for dijet (and possibly trijet production) have recently become available~\cite{Jin:2019dho,Bell:2023yso,Agarwal:2024gws}.

Non-local subtraction methods based on observables sensitive to the transverse momentum $k_T$ of the radiation constitute an alternative path towards higher-order accuracy.
In particular, the transverse momentum $q_T$~\cite{Catani:2007vq} is a well established and extremely successful variable to reach NNLO (and even next-to-next-to-next-to-leading order (N$^3$LO)~\cite{Cieri:2018oms,Camarda:2021ict,Billis:2021ecs,Neumann:2022lft,Chen:2022cgv,Chen:2022lwc}) accuracy in colour-singlet~\cite{Catani:2009sm,Ferrera:2011bk,Catani:2011qz,Ferrera:2014lca,Cascioli:2014yka,Gehrmann:2014fva,Grazzini:2015nwa,Grazzini:2016swo,Grazzini:2017mhc,Kallweit:2020gcp,Garbarino:2025bfg} and heavy-quark hadroproduction \cite{Catani:2019iny,Catani:2020kkl} as well as related processes~\cite{Catani:2022mfv,Buonocore:2022pqq,Buonocore:2023ljm,Buonocore:2025fqs}.
Recently, the transverse momentum of the leading jet has also been considered in this context \cite{Abreu:2022zgo}.
Transverse-momentum--like variables typically lead to small power-suppressed contributions (see e.g. Refs.~\cite{Grazzini:2017mhc,Campbell:2022gdq,Buonocore:2023mne}), and are known to be only mildly sensitive to hadronisation and the underlying event~\cite{Banfi:2010xy,Buonocore:2022mle}.
Attempts to extend $q_T$-slicing to processes with final-state jets, however, have started only in the past few years~\cite{Buonocore:2022mle,Buonocore:2023rdw,Fu:2024fgj}, and initially focused on a suitable definition of the observables and studies at NLO accuracy.
The main challenges to reach NNLO accuracy and beyond are the knowledge of the factorisation properties of multijet cross sections and the computation of the relevant perturbative ingredients.
Recently, some of us started to address these challenges by establishing a factorisation framework for multijet processes in Ref.~\cite{Haag:2025ywj}, and by presenting a general method for the computation of the NNLO quark jet function for $k_T$-like variables in Ref.~\cite{Buonocore:2025ysd}.

In this paper, we build upon the insight obtained in our previous works to construct a new non-local subtraction scheme suitable for NNLO computations with final-state jets.
As a first application, we consider dijet production in $e^+ e^-$ collisions and in $H\rightarrow b \bar b$ decays in NNLO QCD.
Both processes are known fully differentially at N$^3$LO in QCD~\cite{Mondini:2019gid,Chen:2025kez}, and constitute an ideal benchmark for a new subtraction scheme.
Thanks to our recent computation of the NNLO quark jet function and the availability of the NNLO hard functions~\cite{Matsuura:1988sm,Ravindran:2006cg,Anastasiou:2011qx}, most of the theoretical ingredients for NNLO accuracy are available.
The notable exceptions are the NNLO soft function and a possible contribution that breaks the factorised form of the cross section into hard, jet and soft functions. The computation of the above ingredients is one of the main results of this paper.

The paper is organised as follows. In Sect.~\ref{sec:computation} we discuss the radiative functions entering our computation. In Sect.~\ref{sec:soft} we report on the computation of the subtracted soft function, and in Sect.~\ref{sec:factbreak} we evaluate the factorisaton-breaking term that starts contributing at NNLO. The quark jet function was already discussed in Ref.~\cite{Buonocore:2025ysd}. In Sect.~\ref{sec:results} we show our numerical results for the NNLO corrections to \mbox{$e^+e^-\to \gamma^*\to 2$ jets} and \mbox{$\hbb$} and compare the results obtained with different slicing variables.
In Sect.~\ref{sec:summa} we summarise our results. Technical details are given in three Appendices.

\section{The computation}
\label{sec:computation}

We consider the process in which jets are produced through the production and decay of a colourless particle of squared invariant mass $Q^2$. In particular, we will address both \mbox{$e^+e^-\to \gamma^*\to q{\bar q}$} and \mbox{$\mu^+\mu^-\to H\to b{\bar b}$} (with massless bottom quarks).
Ultraviolet (UV) and infrared (IR) divergences are regularised by using conventional
dimensional regularisation in \mbox{$d=4-2\ep$} space-time dimensions. The
$\mathrm{SU}(N_c)$ QCD colour factors are \mbox{$C_F=(N_c^2-1)/(2N_c)$}, \mbox{$C_A = N_c$}, \mbox{$T_R = 1/2$}, and we consider $n_f$ massless quark flavours.
To compute QCD radiative corrections, we use a dimensionful $k_T$-like resolution variable $\qf$ that vanishes in the two-jet limit, and employ it as a slicing variable. As is customary, we introduce a slicing parameter $\qfcut$ and write the (N)NLO cross section as
\begin{equation}
  \label{eq:xs}
d\sigma^{\rm (N)NLO}=d\sigma^{\rm (N)NLO}_{\qf>\qfcut}+d\sigma^{\rm (N)NLO}_{\qf<\qfcut}.
\end{equation}
The evaluation of the above-cut cross section \mbox{$d\sigma _{\qf > \qfcut}$} at (N)NLO requires only an (N)LO calculation and will be discussed in Sect.~\ref{sec:results}.
In the limit \mbox{$\qfcut/\sqrt{Q^2}\ll 1$} the below-cut cross section \mbox{$d\sigma_{\qf < \qfcut}$} can be written as
\begin{equation}
  \label{eq:belowcut}
    d\sigma _{\qf < \qfcut} = d\sigma_{\rm B}\bigg[H\, \mR (\qfcut) \bigg]_{\ep\to 0}\,,
\end{equation}
where $d\sigma_{\rm B}$ is the Born cross section, $H$ denotes the hard function and $\mR(\qfcut)$ can be written as
\begin{equation}\label{eq:mRmaster}
    \mR(\qfcut) = \mJ _{N,q_1} (\qfcut) \mJ _{N,\bar{q}_2} (\qfcut) \mSsub(\qfcut) + \mJS_{N,q_1}(\qfcut) + \mJS_{N,\bar{q}_2}(\qfcut)+{\cal O}(\as^3)\, .
\end{equation}
The dependence on the dimensional-regularisation parameter \mbox{$\ep=(4-d)/2$} in the various radiative functions is always understood: the subscript \mbox{$\ep\to 0$} in Eq.~(\ref{eq:belowcut}) means that
the limit \mbox{$\ep\to 0$} has to be taken at the end. 
The functions $\mJ_{N,q_i}(\qfcut)$, \mbox{$i=1,2$} are the quark jet functions computed by regularising rapidity divergences with the $z_N$ prescription~\cite{Catani:2022sgr}, as discussed in Ref.~\cite{Buonocore:2025ysd}, $\mSsub(\qfcut)$ is the corresponding subtracted soft function, while $\mJS_{N,q_i}(\qfcut)$ is a factorisation-breaking contribution, which appears, for example, when the $\qf$ variable is defined through the $k_{T}$ jet clustering algorithm and the $E$-scheme is employed in the recombination of protojets~\cite{Haag:2025ywj}. The perturbative expansions of the hard, jet and soft functions read\footnote{The dependence on $\qfcut$, $Q^2$ and $\mu$ in the expansion coefficients is left understood.}
\begin{align}
  \label{eq:exp}
H &=1+\frac{\alpha_0}{\pi} \left(\frac{\mu^2}{Q^2}\right)^{\ep} H^{(1)}+\left(\frac{\alpha_0}{\pi}\right)^2 \left(\frac{\mu^2}{Q^2}\right)^{2\ep} H^{(2)}+{\cal O}\left(\alpha_0^3\right)\,,\nonumber \\
\mJ _{N,q_1} (\qfcut) &= 1 + \frac{\azero}{\pi} \left( \frac{\mu^2}{\qfcut^2} \right)^{\epsilon} \mJ_{N,q_1}^{(1)} + \left( \frac{\azero}{\pi} \right)^2 \left(\frac{\mu^2}{\qfcut^2}\right)^{2\epsilon} \mJ^{(2)}_{N,q_1} + \mathcal{O}(\alpha_{0}^3)\,, \\
\mSsub(\qfcut) &= 1 + \frac{\as(\mu)}{\pi} \left(\frac{\mu^2}{\qfcut^2}\right)^{\epsilon} \mS_{\rm sub}^{(1)} + \left( \frac{\as(\mu)}{\pi} \right)^2 \left(\frac{\mu^2}{\qfcut^2}\right)^{2\epsilon} \mS_{\rm sub}^{(2)} + \mathcal{O}(\as ^3)\,. \nonumber
\end{align}
The factorisation-breaking term starts to contribute at NNLO,
\begin{equation}
\mJS_{N,q_1}(\qfcut) = \left(\frac{\as}{\pi}\right)^2 \left( \frac{\mu^2}{\qfcut^2} \right)^{2\epsilon} \mJS_{N,q_1}^{(2)}+ \mathcal{O}(\as^3).
\end{equation}
The bare coupling $\alpha_0=\alpha_0(\mu)$ in Eq.~(\ref{eq:exp}) is related to the (dimensionless) unrenormalised coupling $g_{\mathrm S}^u$ in the QCD Lagrangian by
\begin{equation}
\as^u\mu_0^{2\ep} S_\ep=\alpha_0(\mu) \mu^{2\ep} \,, 
\end{equation}
where \mbox{$\asbare=(g_{\mathrm S}^u)^2/(4\pi)$}
and $S_\ep$ is the customary spherical factor \mbox{$S_\ep=(4\pi)^\ep e^{-\ep \gamma_E}$}.
The relation between $\azero$ and the renormalised coupling $\as(\mu)$ in the $\MSbar$ scheme is given by
\begin{equation}
\azero = \as(\mu) Z_\alpha \,, ~~~~~~~~~~~Z_\alpha = 1-\frac{\as (\mu)}{\pi} \frac{\beta_0}{\epsilon} + \mathcal{O}(\as^2)  \,, 
\end{equation}
where \mbox{$\beta_0=(11C_A-4n_fT_R)/12$}. In the case of \mbox{$\hbb$} the Yukawa coupling $y_b$ is renormalised in the $\MSbar$ scheme at the scale $\mu$ and the Born level cross section $d\sigma_{\rm B}$ is proportional to $y^2_b(\mu)$. The subtracted soft function is more conveniently expanded in the renormalised coupling $\as(\mu)$ since, as we will see below, it is finite as \mbox{$\ep\to 0$} in our implementation of the $z_{N}$ prescription.

In the following, we discuss the various radiative functions in turn.
We start from the hard function, which does not depend on the definition of the $\qf$ variable, but is the only truly process-dependent ingredient of the computation. The explicit expressions of the coefficients $H^{(1)}$ and $H^{(2)}$ can be obtained from the one- and two-loop \mbox{$\gamma^*\to q{\bar q}$}~\cite{Matsuura:1988sm} and \mbox{$\hbb$}~\cite{Ravindran:2006cg} virtual amplitudes, and are reported in Appendix \ref{app:hard}.

The other perturbative ingredients depend on the definition of the
  slicing variable. To be definite, we focus on a specific choice of the $\qf$
  variable, a variant of the Durham $y_{23}$ jet resolution parameter~\cite{Catani:1991hj} which is particularly suitable to handle final-state radiation. We
  define the $\qf$ variable by using a recursive recombination jet algorithm
  with the distance
\begin{equation}\label{eq:dijdef}
  d_{ij}^2 = \frac{E_i^2 E_j^2}{(E_i+E_j)^2} 2(1-\cos \vartheta _{ij})\, ,
\end{equation}
as employed in Ref.~\cite{Buonocore:2025ysd}. This particular choice simplifies the calculation
of the jet function. Moreover, the choice of the angular distance
\mbox{$2(1-\cos \vartheta _{ij})$} renders the subtracted soft function independent of the
Born kinematics.

We run the recursive recombination on a generic $n+k$ parton system (\mbox{$n=2$} in our case) until \mbox{$n+1$}
pseudo-partons are left. On this configuration, the distances $d_{ij}$ are calculated again, and we define \mbox{$\qf=\min\{d_{ij}\}$}. In the following, two recombination schemes are considered:
\begin{itemize}
    \item $E$-scheme, where the momentum $k_{ij}$ obtained by recombining particles with momenta $k_i$ and $k_j$ is
    \begin{equation}
        k_{ij} = k_i + k_j\, .
    \end{equation}
    \item Winner-take-all (WTA) scheme~\cite{Bertolini:2013iqa}, where  the momentum of the recombined particle is
    \begin{equation}
        k_{ij} = (E_i+E_j)\left(\frac{k_i}{E_i}\theta(E_i-E_j) + \frac{k_j}{E_j} \theta (E_j-E_i)\right)\, .
    \end{equation}
\end{itemize}
In the $E$-scheme, the recombined momentum is simply the sum of the original
momenta, and is in general massive. In the WTA scheme, the recombined momentum
is massless, and has the direction of the more energetic pseudo-parton.

For a single emission the
variable defined above approaches the transverse momentum relative to the jet
direction in the two-parton collinear limit. For such variables, the computation of the jet function has been presented in Ref.~\cite{Buonocore:2025ysd}. In this work we set the auxiliary
reference time-like vector $N$ used in the $z_N$ prescription to \mbox{$N^{\mu}=Q^{\mu}$}, where $Q^{\mu}$ is the total incoming momentum. Hence,
the NLO coefficient $\mJ^{(1)}_{N,q}$ reads
\begin{equation}
    \mJ^{(1)}_{N,q} = \mJ^{(1)}_{N,\bar{q}} =  \frac{e^{\epsilon \gamma_E}}{\Gamma(1-\eps)} C_F \left[ \frac{1}{2\epsilon^2} + \frac{L}{\eps} + \frac{3}{4\eps} + \frac{1}{4} \right] \,,
\end{equation}
while at NNLO we have
\begin{equation}
 {\J}_{N,q}^{(2)} = \mJ^{(2)}_{N,\bar{q}} = L^2 \sum_{k=0}^2 \frac{D_k}{\eps^k} + L \sum _{k=0}^3 \frac{A_k}{\eps ^k} + \sum _{k=0}^4 \frac{B_k}{\eps ^k}\, ,
\end{equation}
where \mbox{$L= \ln\left(\qfcut/\sqrt{Q^{2}}\right)$}, and the coefficients $D_k$, $A_k$ and $B_k$ are given in Eqs.~(51)--(53) of Ref.~\cite{Buonocore:2025ysd}.

\subsection{The subtracted soft function}
\label{sec:soft}

We now come to the soft function. As discussed in
Refs.~\cite{Buonocore:2025ysd,Haag:2025ywj}, the regularisation of rapidity
divergences we are using is such that the soft endpoint of the collinear
splitting kernels is no longer scaleless. As a result, a non-vanishing zero-bin
contribution is generated, which needs to be subtracted from the soft function
to avoid double counting. This leads to the definition of the
subtracted soft function $\mS_{\rm sub}$ entering Eq.~(\ref{eq:mRmaster}). In the
following, we report the soft integrals including the zero-bin contributions
required for the calculation of $\mS_{\rm sub}$ up to NNLO for the processes we are considering. We label the
momenta of the hard quark and anti-quark legs with $p_{1}$ and $p_{2}$,
respectively, while we use $k$ and $l$ to denote the momenta of up to two
additional soft partons. In the soft limit, momentum conservation implies
\mbox{$Q = p_1+p_2$}. 

At NLO, the subtracted soft function is defined as
\begin{equation}\label{eq:SsubNLO}
  \mS_{\rm sub}^{(1)} = \qf_{\rm cut}^{2\eps} \frac{e^{\eps \gamma_E}}{\Gamma(1-\eps)}\int \frac{d^dk}{\Omega_{d-2}} \delta _+(k^2) \left[\JJ_g^{(0)}(k) \theta (\qfcut - \qf_S)   - \sum _{i=1}^2  2C_F \frac{p_1\cdot p_{2}}{(p_i\cdot k)( Q\cdot k)}\theta (\qfcut - \qf _{C_i,S}) \right] \,,
\end{equation}
where the soft factor $\JJ_g^{(0)}(k)$ reads
\begin{equation}
  \label{eq:jj0}
  \JJ_g^{(0)}(k)=2C_F \frac{p_1\cdot p_2}{(p_1\cdot k)(p_2\cdot k)}\,.
\end{equation}
In Eq.~\eqref{eq:SsubNLO}, the soft factor is integrated over the soft phase space and the soft-collinear (zero-bin) contribution is subtracted. The variables $\qf_S$ and $\qf_{C_i,S}$ are the soft and
soft-collinear limits of the $\qf$ variable, respectively. For the variables
considered in this paper, their expressions in the partonic centre-of-mass frame
are
\begin{equation}\label{eq:limitsNLO}
  \qf_{C_i,S}=k_\perp\,,~~~~~~~~\qf^2_{S}=2k_0^2(1-|\cos\vartheta|)\,,
\end{equation}
where $\vartheta$ is the angle between $k$ and $p_1$ and $k_{\perp}$ is the
transverse momentum with respect to the quark.
Using Eq.~\eqref{eq:limitsNLO}, we can also write $\mS_{\rm sub}^{(1)}$ as
\begin{equation}
  \mS_{\rm sub}^{(1)} = \qf_{\rm cut}^{2\eps} \frac{e^{\eps \gamma_E}}{\Gamma(1-\eps)}2C_F\int \frac{d^dk}{\Omega_{d-2}} \delta _+(k^2)  \frac{p_1\cdot p_2}{(p_1\cdot k)(p_2\cdot k)}
  \left[\theta (\qfcut - \qf_S)   - \theta (\qfcut - k_\perp) \right] \,,
\end{equation}
and we find\footnote{The subtracted soft function generally depends on the
  momenta of the hard configuration even for dijet production. This is not the case for the variables considered in this work thanks to the form of the angular distance in Eq.~\eqref{eq:dijdef}.}
\begin{equation}
\mS_{\rm sub}^{(1)} =   -C_F \left[ \frac{\pi^2}{12} + \frac{\zeta_3}{8} \eps + \mathcal{O}(\eps^2) \right]\, .
\end{equation}
At NNLO the subtracted soft function receives contributions from soft-gluon emission at one-loop order~\cite{Catani:2000pi}, from double soft-gluon emission and from soft quark--antiquark emission at tree level~\cite{Catani:1999ss}. We start from the contribution arising from soft-gluon emission at one-loop order. Given the trivial colour structure of the Born-level process, the relevant soft-factorisation formula is obtained through a simple rescaling of the tree-level eikonal factor.
The corresponding subtracted soft function is
\begin{align}\label{eq:S2gsub}
    \mS^{(2)}_{g,\rm sub} & = -\qfcut^{4\eps} c_S C_A C_F \left( \frac{e^{\eps \gamma_E}}{\Gamma(1-\eps)} \right)\frac{\cos (\pi \eps)}{\eps^2} \int \frac{d^dk}{\Omega_{2-2\eps}}\delta_+(k^2) k_\perp^{-2\eps} \nonumber \\
    & \times \frac{p_1\cdot p_2}{(p_1\cdot k)(p_2\cdot k)}\big( \theta(\qfcut-\qf_S) - \theta(\qfcut-k_\perp)\big)\,,
\end{align}
where $c_S$ is given by
\begin{equation}
  \label{eq_cs_definition}
  c_S=\frac{e^{\epsilon\gamma_E}\Gamma^3(1-\epsilon) \Gamma^2(1+\epsilon)}{\Gamma(1-2 \epsilon)}=1 +\frac{\pi ^2 }{12}\epsilon ^2-\frac{7 \zeta_3 }{3}\epsilon ^3-\frac{13 \pi ^4}{480} \epsilon ^4+ \mathcal{O}(\epsilon^5)\, .
\end{equation}
We find
\begin{align}
  \mS^{(2)}_{g,\rm sub} & =C_A C_F \left( \frac{\pi ^2}{24 \epsilon ^2}+\frac{\zeta_3}{8 \epsilon } -\frac{7}{4} \zeta_3 \ln 2+\frac{\pi ^4}{720}-\frac{1}{12} \ln ^4 2+\frac{1}{12} \pi ^2 \ln ^2 2-2 \operatorname{Li}_4\!\left(\frac{1}{2}\right)+\mathcal{O}(\epsilon)\right)\, .
\end{align}
We now consider the double-real emission. The corresponding contribution to the subtracted soft function has been presented and discussed in the general case in Ref.~\cite{Haag:2025ywj}.
 In the following, we report the relevant expressions
  for the case at hand, after performing the colour algebra.

The contribution from the double-soft $q\bar{q}$ radiation is
\begin{align}
  \label{eq:softqq}
    \mathcal{S}^{(2)}_{q\bar{q},\rm sub} & = \qfcut^{4\eps} \left(\frac{e^{\eps \gamma_E}}{\Gamma(1-2\eps)}\right)^2 2C_F n_f T_R \int \frac{d^dk}{\Omega_{d-2}}\delta_+(k^2)\int \frac{d^dl}{\Omega_{d-2}}\delta _+(l^2) \mathcal{S}_{q\bar{q}}(p_1,p_2;k,l) \nonumber \\
    & \times \bigg[\frac{p_1\cdot p_2}{p_1\cdot(k+l)\, p_2\cdot (k+l)}\Theta_S- \bigg( \frac{p_1\cdot p_2}{p_1\cdot(k+l)\, Q\cdot(k+l)}\Theta_{C_1,S} + (1\leftrightarrow2) \bigg)\bigg] \,,
\end{align}
where the soft factor $\mS_{q\bar{q}}(p_1,p_2;k,l)$ reads
\begin{equation}
  \label{eq:softfacqq}
    \mS_{q\bar{q}}(p_1,p_2;k,l) = \frac{1}{k\cdot l}\left(1-\frac{[(p_1\cdot l)(p_2\cdot k) - (p_1\cdot k)(p_2\cdot l)]^2}{(p_1\cdot p_2)(k\cdot l)\left[p_1\cdot(k+l)\right]\left[p_2\cdot(k+l)\right]}\right)\,.
\end{equation}
The first term in the square bracket in Eq.~(\ref{eq:softqq}) represents the double-soft $q\bar{q}$ contribution, while the second term is the soft-collinear (zero-bin) subtraction. Here and in the following we use a shorthand notation for the $\Theta$ functions,
\begin{equation}
\Theta_A\equiv \theta(\qfcut-\qf_A)\,,
\end{equation}
where $\qf_A$ is the approximation of $\qf$ in the limit $A$. In case of
subtracted contributions, $\qf_{A,B}$ is the approximation of $\qf$ in which we
first expand in the limit $A$, and then in the limit $B$, and similarly for
$\qf_{A,B,C}$. The relevant expressions in the case of our variable are given in Appendix \ref{app:limits}.

The double soft-gluon emission is divided into an abelian and a non-abelian contribution. The non-abelian contribution reads
\begin{align}
\label{eq: softggnab}
    \mathcal{S}^{(2),(\rm nab)}_{gg,\rm sub} & = \qfcut^{4\eps} \left( \frac{e^{\eps \gamma_E}}{\Gamma(1-\eps)} \right)^2 C_F C_A \int \frac{d^dk}{\Omega_{d-2}}\delta_+(k^2)\int \frac{d^dl}{\Omega_{d-2}}\delta_+(l^2) \mS_{gg}(p_1,p_2;k,l)  \nonumber \\
    & \times \bigg[ \frac{p_1\cdot p_2}{p_1\cdot(k+l)\,p_2\cdot(k+l)}\Theta_S -\bigg(\frac{p_1\cdot p_2}{p_1\cdot(k+l)\,Q\cdot(k+l)} \Theta_{C_1,S} + (1\leftrightarrow 2) \bigg) \bigg] \, ,
\end{align}
where the soft factor $\mS_{gg}(p_i,p_j;q_1,q_2)$ is defined as\footnote{In Eq.~(\ref{eq:softfacqq}) and (\ref{eq:softfacgg}) gauge invariance has been used to redefine the soft factors of Ref.~\cite{Catani:1999ss} to make them vanish as $i=j$.}
\begin{align}
  \label{eq:softfacgg}
    & \mS_{gg}(p_i,p_j;q_1,q_2) = \frac{(1-\eps)[(p_i\cdot q_2)(p_j \cdot q_1) - (p_i \cdot q_1)(p_j \cdot q_2)]^2}{(p_i \cdot p_j)(q_1\cdot \
q_2)^2 p_i \cdot (q_1 + q_2) p_j \cdot (q_1 + q_2)} \nonumber \\
    & + \frac{1}{2q_1 \cdot q_2} \bigg[ \frac{p_i \cdot (q_1 + q_2) p_j \cdot (q_1 + q_2)}{(p_i \cdot q_2)(p_j \cdot q_1)} + \frac{p_i \cdot (\
q_1 + q_2) p_j \cdot (q_1 + q_2)}{(p_i \cdot q_1)(p_j \cdot q_2)} \nonumber \\
    &  + \frac{p_i \cdot q_1}{p_i \cdot q_1} + \frac{p_i \cdot q_2}{p_i \cdot q_1} + \frac{p_j \cdot q_2}{p_j \cdot q_1} + \frac{p_j \cdot q_1\
}{p_j \cdot q_2} - 4 \bigg] \nonumber \\
    & - \frac{(p_i \cdot p_j)[p_i \cdot (q_1 + q_2) p_j \cdot (q_1 + q_2) + (p_i \cdot q_1)(p_j \cdot q_1) + (p_i \cdot q_2)(p_j \cdot q_2)]}{\
2(p_i \cdot q_1)(p_i \cdot q_2)(p_j \cdot q_1)(p_j \cdot q_2)} \, ,
\end{align}
and the soft-collinear subtractions are analogous to those in Eq.~(\ref{eq:softqq}).

The case of the abelian double-soft gluon contribution is more difficult to deal with. In this case, the implementation of the $z_N$ prescription to evaluate the collinear contributions is not straightforward~\cite{Buonocore:2025ysd}, and, correspondingly, additional subtractions are needed~\cite{Haag:2025ywj}.
The abelian double-soft gluon contribution can eventually be rewritten as
\begin{align}
  \label{eq:dsoftnab}
     \mS^{(2),(\rm ab)}_{gg,\rm sub} &= \qfcut^{4\eps} \left( \frac{e^{\eps \gamma_E}}{\Gamma(1-\eps)} \right)^2 2C_F^2 \int \frac{d^dk}{\Omega_{d-2}}\delta_+(k^2) \int \frac{d^dl}{\Omega_{d-2}}\delta_+(l^2) \nonumber \\
    & \times\bigg\{ \frac{(p_1\cdot p_2)^2}{p_1\cdot k \,\, p_2\cdot k \,\, p_1\cdot l \,\, p_2\cdot l} (\Theta_S - \Theta_{\rm max}) \nonumber \\
    & - \bigg[ \frac{(p_1\cdot p_2)^2}{p_1\cdot k \,\, Q\cdot k\,\, p_1\cdot l \,\,Q\cdot l} (\Theta _{C_1,S} - \Theta_{C_1,SC_1,S} - \Theta_{C_1,C_1S,S} + \Theta _{\rm max}) + (1\leftrightarrow 2)\bigg] \nonumber \\
    & - \bigg[\bigg(\frac{(p_1\cdot p_2)^2}{p_1\cdot k\,Q\cdot k \, p_1\cdot l \, p_2\cdot l}(\Theta_{C_1S,S} - \Theta_{\rm max}) + (k\leftrightarrow l) \bigg) + (1\leftrightarrow 2)\bigg] \bigg\}\, ,
\end{align} 
where we have also defined
\mbox{$\Theta_{\rm max}\equiv\theta(\qfcut-{\rm max}(l_{\perp},k_{\perp}))$}.
Adding all the contributions, together with the term arising from the renormalisation of the
strong coupling, the $\epsilon$ poles cancel out and the final result for the NNLO subtracted soft function can be written as
\begin{align}
  \mS_{\rm sub}^{(2)}
  &= \mS^{(2)}_{g,\rm sub} +  \mathcal{S}^{(2)}_{q\bar{q},\rm sub} + \mathcal{S}^{(2),(\rm nab)}_{gg,\rm sub}  + \mS^{(2),(\rm ab)}_{gg,\rm sub}  -  \frac{\beta_0}{\eps}\mS_{\rm sub}^{(1)} \left(\frac{\qfcut}{\mu}\right)^{2\ep}\notag \\
  &= C_{F} \beta_{0} \frac{\pi^{2}}{6} \ln\frac{\qfcut}{\mu} + C_FC_A \mS_{A}  + C_F n_f T_R \mS_f + C_F^2 \mS_{2F} + \mathcal{O}(\epsilon)\, ,
\end{align}
where the coefficients in the $E$-scheme are given by
\begin{equation}
    \mS_{2F} = -10.068(4), \quad \mS_{A} = -1.3583(4), \quad \mS_f = 0.25370(9) \, ,
\end{equation}
while in the WTA scheme they are
\begin{equation}
    \mS_{2F} = -8.1900(8) , \quad \mS_A = -1.43131(1), \quad \mS_f = 0.265098(3) \, .
\end{equation}

For the computation of the integrals in Eqs.~\eqref{eq:S2gsub}--\eqref{eq:dsoftnab}, we proceed along
the lines of Refs.~\cite{Bell:2018oqa,Bell:2020yzz}. More specifically, we adopt the phase space
parametrization
\begin{equation}
    m_T = \sqrt{(k^+ + l^+)(k^-+l^-)}, \quad y = \frac{k^+ + l^+}{k^- + l ^-}, \quad a = \sqrt{\frac{k^- l^+}{k^+ l^-}}, \quad b = \sqrt{\frac{k^- k^+}{l^- l^+}}, \quad \cos \varphi = \frac{\vec{k}_{\perp} \cdot \vec{l}_{\perp}}{k_{\perp} l_{\perp}} \, ,
  \end{equation}
  where $+,-,\perp$ denote light-cone components of $k$ and $l$ with respect to
  $p_{1,2}$. The integration over the dimensionful variable $m_{T}$ can be
  performed analytically. The remaining integrals are performed numerically,
  taking care of overlapping singularities using either sector decomposition or
  a suitable change of variables. All numerical coefficients are computed with a dedicated Fortran code using {\sc Cuba}~\cite{Hahn:2004fe}.

\subsection{Factorisation-breaking contribution}
\label{sec:factbreak}

Our study of the \mbox{$\qf\to 0$} limit is based on an expansion in the relevant hard, collinear, and soft
regions. Mixing between different regions can arise at higher-loop orders. If
the observable under consideration satisfies a factorisation theorem, the
structure of such mixing terms is fully determined: at a given loop order, they
can be expressed in terms of products of lower-order ingredients.

For the class of transverse-momentum observables considered in this work, mixing
terms between the collinear and soft sector --- beyond those predicted by a
factorisation ansatz --- may first appear at $\mathcal{O}(\as^{2})$.
Physically, this originates from the definition of the variable via a clustering
procedure with a specific prescription for the treatment of momentum recoil at
each merging step of two nearby partons. In a configuration with multiple
collinear emissions, the proto-jet axis recoils after each merging step, until
only a single parton and a proto-jet remain. The observable is then defined as
the transverse momentum of the parton with respect to the axis of the final
proto-jet. If the jet-axis recoil is affected at leading power by soft
emissions, factorisation-violating terms arise. An example of a merging prescription
insensitive to soft recoil is the WTA scheme. Consequently,
the variant of the variable defined with the WTA scheme satisfies a factorisation theorem~\cite{Haag:2025ywj}, in close analogy to the observables studied in
Refs.~\cite{Chien:2020hzh,Chien:2022wiq,Fu:2024fgj}.

The situation is different for the $E$-scheme. Here, the recoil is sensitive to
soft emissions already at leading power, which induces a breaking of
factorisation in cumulant space. In what follows, we focus exclusively on collinear and soft
sectors. Factorisation in cumulant space requires, roughly speaking, that the
observable behaves as the maximum of its collinear and soft limits.

To be precise, in a region of phase space where the set of final-state partons with momenta $\{k_i\}$ can be divided into three subsets, one collinear to the hard quark ($\{k_i\}_{C_1}$), one collinear to the hard anti-quark ($\{k_i\}_{C_2}$), and one soft ($\{k_i\}_S$), the observable must satisfy\footnote{Note that transverse-momentum conservation in the collinear sectors is affected by the soft particles. Parametrizing the collinear momenta in sector $C_1$ as
\mbox{$k_i = z_i p_1 + k_{i,\perp}-\frac{k_{i,\perp}^2}{2z_i\ p_1\cdot p_2}p_2$}, to achieve factorisation, the function $\qf_{C_1}$ is only allowed to depend on the momentum fractions $z_i$ and the recoil-independent transverse momenta \mbox{$\widetilde{k}_{\perp i}^\mu=k_{\perp i}^\mu-\frac{z_i}{\sum_{k=1}^n z_k} \sum_{j=1}^n k_{\perp j}^\mu$}. The same holds for $\qf_{C_2}$. A more detailed discussion can be found in Ref.~\cite{Haag:2025ywj}.}
\begin{equation}
  \label{eq:maximum_factorisation}
\qf(\{k_i\}) = \max\{\qf_{C_1}(\{k_i\}_{C_1}),\qf_{C_2}(\{k_i\}_{C_2}),\qf_{S}(\{k_i\}_{S})\}\,.
\end{equation}

At $\mathcal{O}(\as^{2})$, one must consider configurations with three
partons: two collinear to leg $i$ (\mbox{$i=1,2$}) with momenta $k_{1},k_{2}$, and one soft
parton with momentum $k_{3}$. The leading-power contribution to the cross
section in Eq.~\eqref{eq:xs} for such configurations is given by
\begin{align}\label{eq:fbmaster}
  \qf_{\rm cut}^{4\eps} \left( \frac{e^{\eps \gamma_E}}{\Gamma(1-\eps)} \right) ^2\int _0^1 & dz \frac{d^{d-2}\vec{k}_{1,\perp}}{\Omega_{d-2}}\frac{\hat{P}^{(0)}_{N,qg}(z)}{k_{1,\perp}^2} \int \frac{d^dk_3}{\Omega_{d-2}} \delta_+(k_3^2) \nonumber \\
  & \times \bigg\{ \JJ_{g}^{(0)}(k_3) \theta (\qfcut - \qf_{C_iS}(\{k_1,k_2\},\{k_3\}))  \nonumber\\
  & - 2C_F \frac{p_1 \cdot p_2}{(p_{\bar{\imath}} \cdot k_{3})(Q\cdot k_3)}  \theta (\qfcut - \qf_{C_i}(\{k_1,k_2\})) \theta (\qfcut - \qf_{C_{\bar{\imath}},S}(k_3))  \nonumber\\
    & - 2C_F \frac{p_1 \cdot p_2}{(p_i \cdot k_3)(Q\cdot k_{3})} \theta (\qfcut - \qf_{C_i,C_iS}(\{k_1,k_2\},\{k_3\}))  \bigg\}\,.
\end{align}
The soft-collinear contribution in Eq.~(\ref{eq:fbmaster}) is driven by the tree-level splitting kernel $\hat{P}^{(0)}_{N,qg}(z)$ (we use the notation of Ref.~\cite{Buonocore:2025ysd}) and by the tree-level squared-soft current $ \JJ_{g}^{(0)}(k_3)$ of Eq.~(\ref{eq:jj0}), with
the subtraction of the zero-bin contributions arising from the double-collinear limit
(\mbox{$k_{1}\parallel k_{2}$} and \mbox{$k_{3}\parallel p_{\bar{\imath}}$}, where $\bar{\imath}$
denotes the non-$i$ leg), and from the triple-collinear limit
(\mbox{$k_{1}\parallel k_{2}\parallel k_{3}$}), shown on the third and fourth
lines, respectively. We observe a potentially non-factorising interplay of collinear
and soft momenta in the theta functions of the second and fourth lines, while in
the double-collinear zero-bin case the observable factorises trivially.

As discussed, if the observable satisfies
\begin{align}\label{eq:fb-maxdef}
\qf_{C_{i}S} (\{k_1,k_2\},\{k_3\}) &= \max\{\qf_{C_{i}}(k_{1},k_{2}),\qf_{S}(k_{3})\}\,, \nonumber \\
\qf_{C_{i},C_{i}S} (\{k_1,k_2\},\{k_3\}) &= \max\{\qf_{C_{i}}(k_{1},k_{2}),\qf_{C_{i},S}(k_{3})\} \,,
\end{align}
the integral in Eq.~\eqref{eq:fbmaster} reduces to the product of the ${\cal O}(\as)$ jet and subtracted soft functions,
$\mJ_{N,i}^{(1)} \mS_{\rm sub}^{(1)}$, as predicted by cumulant factorisation. It is therefore natural to decompose the calculation into
the contribution that factorises and the remainder that does not, by adding and
subtracting the theta functions in Eq.~\eqref{eq:fb-maxdef}. The
factorisation-breaking contribution then reads~\cite{Haag:2025ywj}
\begin{align}\label{eq:fbmaster-2}
    \mJS_{N,i}^{(2)} & = \qf_{\rm cut}^{4\eps} \left( \frac{e^{\eps \gamma_E}}{\Gamma(1-\eps)} \right) ^2\int _0^1 dz \frac{d^{d-2}\vec{k}_{1,\perp}}{\Omega_{d-2}}\frac{\hat{P}^{(0)}_{N,qg}(z)}{k_{1,\perp}^2} \int \frac{d^dk_3}{\Omega_{d-2}} \delta_+(k_3^2) \nonumber \\
    & \times \bigg\{ \JJ^{(0)}(k_3) \bigg[ \theta (\qfcut - \qf_{C_iS}(\{k_1,k_2\},\{k_3\})) - \theta (\qfcut - \max (k_{1,\perp},\qf_S(k_3))) \bigg] \nonumber \\
    & -2C_F \frac{p_1 \cdot p_2}{(p_i \cdot k_3)(Q\cdot k_{3})}\bigg[ \theta (\qfcut - \qf_{C_i,C_iS}(\{k_1,k_2\},\{k_3\})) - \theta (\qfcut - \max (k_{1,\perp},k_{3,\perp})) \bigg] \bigg\} \nonumber \\
    & = A_{\rm FB}^{(0)} L + B_{\rm FB} ^{(0)}\;,
\end{align}
where we used that the resolution variable in the presence of a single collinear emission is
\begin{equation}
  \label{eq:qC_NLO}
   \qf_{C_{i}}(k_{1},k_{2})=k_{1,\perp}=k_{2,\perp}\, .
\end{equation}
The factorisation-breaking term $\mJS_{N,i}^{(2)}$ is enhanced by a single
logarithm $L$, and thus contributes to the resummation of the observable at NNLL
accuracy. The integrals in Eq.~\eqref{eq:fbmaster-2} are finite, owing to the
differences in the theta functions and the use of the $z_{N}$ prescription, and can be
evaluated numerically.

When the observable factorises in cumulant space, one has \mbox{$\mJS_{N,i}^{(2)}=0$}, as in the case of
the WTA scheme considered here. By contrast, in the $E$-scheme, factorisation is
broken, leading to a non-vanishing $\mJS_{N,i}^{(2)}$. Our observable is a
variant of the Durham $y_{23}$~\cite{Catani:1991hj}, for which a factorisation theorem (when the $E$-scheme is used) is not known. The $y_{23}$ variable has been resummed at
NNLL accuracy in Ref.~\cite{Banfi:2016zlc} using the numerical resummation
framework provided by the ARES formalism~\cite{Banfi:2014sua}.
The observable-dependent coefficient $A_{\rm FB}^{(0)}$ of the logarithmic term directly corresponds to the analogous term in the fixed-order expansion of the ARES formula and is part of the set of corrections required for the resummation of the desired observable at NNLL.\footnote{We recall that our definition of the distance used in the recombination scheme Eq.~\eqref{eq:dijdef} is different from that used in Ref.~\cite{Banfi:2014sua}.} The constant term instead contributes at N$^{3}$LL
accuracy, representing a novel result of this work.

Our numerical results for the coefficients are
\begin{equation}
A_{\rm FB}^{(0)} = (-0.45446\pm 0.00008)C_F^2, \quad B_{\rm FB}^{(0)} =( -0.81980 \pm 0.00033)C_F^2\,.
\end{equation}
We remark that the WTA scheme is
not the only possible way to construct an observable insensitive to soft recoil.
For instance, the Cambridge $y_{23}$~\cite{Dokshitzer:1997in,Bentvelsen:1998ug,vanBeekveld:2025zjh} is known to factorise to all
orders, a consequence of the use of the Cambridge algorithm to determine the
clustering history. As in Sec.~\ref{sec:soft} the numerical coefficients are computed with a dedicated Fortran code using {\sc Cuba}~\cite{Hahn:2004fe}.

\subsection{Complete results for $d\sigma_{\qf<\qfcut}$}

We are now in the position to assemble the various perturbative ingredients and to construct the complete expression for $d\sigma _{\qf < \qfcut}$.
In the product of the radiative functions in Eq.~(\ref{eq:belowcut}), the poles cancel order by order in $\as$. We can therefore safely take the \mbox{$\ep\to 0$} limit to obtain
\begin{align}
    & d\sigma _{\qf < \qfcut} = d\sigma_{\rm B} \left[ 1+\frac{\as}{\pi} \Sigma^{(1)} + \left(\frac{\as}{\pi}\right)^2 \Sigma ^{(2)}+ \mathcal{O}(\as ^3)\right] \, , \nonumber  \\
    & \Sigma ^{(1)} = \sum _{k=0}^2 \Sigma ^{(1,k)}L^k, \quad \Sigma ^{(2)} = \sum _{k=0}^4 \Sigma ^{(2,k)}L^k \, .
\end{align}

At NLO the unresolved cross section for dijet production is given by\footnote{In this section we limit ourselves to presenting results at scale \mbox{$\mu=Q$}. The general expressions for the coefficients $\Sigma^{(i,j)}$ can be obtained from those presented here by using renormalisation group invariance.}
\begin{equation}
\Sigma^{(1)}_{\twoj}= C_F \left(-2 L^2 -3 L + \frac{5\pi^2}{12} - \frac{7}{2} \right) \, .
\end{equation}
The coefficients $\Sigma ^{(i,j)}$ then read 
\begin{equation}
    \Sigma^{(1,2)} = -2C_F, \quad~~~ \Sigma ^{(1,1)} = -3C_F, \quad~~~ \Sigma ^{(1,0)}_\twoj = C_F \left(\frac{5\pi^2}{12} - \frac{7}{2}\right) \, .
\end{equation}
For the decay \mbox{$\hbb$} we find
\begin{equation}
    \Sigma ^{(1)}_\hbb = C_F \left( -2 L^2 -3L +\frac{5\pi^2}{12}-\frac{1}{2} \right) \, ,
\end{equation}
so the only different coefficient is $\Sigma ^{(1,0)}$, which reads
\begin{equation}
    \Sigma ^{(1,0)}_\hbb = C_F\left(\frac{5\pi^2}{12}-\frac{1}{2}\right) \, .
\end{equation}

We now turn to the NNLO results. By considering the $\mathcal{O}(\as^2)$ contribution to the product of the various radiative functions in Eq.~(\ref{eq:belowcut}), we can verify that all the $\eps$-poles cancel out leaving the finite result
\begin{align}
    \Sigma^{(2)} = \sum _{k=0}^4 \Sigma ^{(2,k)}L^k\, .
\end{align}
The first two coefficients are universal for both processes and both considered recombination schemes and are given by
\begin{align}
      & \Sigma^{(2,4)} = 2C_F^2\,, \nonumber \\
    & \Sigma^{(2,3)} = 6C_F^2 + \frac{22}{9}C_FC_A-\frac{8}{9}C_F n_f T_R\, .
\end{align}
For dijet production in the $E$-scheme, the remaining coefficients read
\begin{align}
    & \Sigma^{(2,2)}_{\twoj,E} = \left(\frac{23}{2} -\pi ^2+\log ^2(2) \right)C_F^2+\left( \frac{\pi^2}{6} - \frac{35}{36} \right)C_FC_A + \frac{1}{9}C_F n_f T_R \, , \nonumber \\
    & \Sigma^{(2.1)}_{\twoj,E} = (-2.4892(3))C_F^2 + (-1.9558(1))C_FC_A + (1.28116(4))C_F n_f T_R \, , \nonumber \\
    & \Sigma^{(2,0)}_{\twoj,E} = (-2.696(2))C_F^2 + (5.935(1))C_FC_A + (-1.6723(2))C_F n_f T_R \, ,
\end{align}
while in the WTA scheme they read
\begin{align}
    & \Sigma^{(2,2)}_{\twoj,{\rm WTA}} = \left( \frac{23}{2}-\frac{5 \pi ^2}{6} \right)C_F^2 + \left( \frac{\pi^2}{6}-\frac{35}{36} \right)C_F C_A + \frac{1}{9}C_F n_f T_R \, , \nonumber \\
    & \Sigma^{(2,1)}_{\twoj,{\rm WTA}} = (2.8056(1))C_F^2 + (-2.84899(4))C_F C_A + (1.3838(2))C_F n_f T_R \, , \nonumber \\
    & \Sigma ^{(2,0)}_{\twoj,{\rm WTA}} = (8.467(1))C_F^2 + (2.9071(7))C_FC_A + (-1.3659(3))C_F n_f T_R \, .
\end{align}
For the \mbox{$\hbb$} decay in the $E$-scheme we have
\begin{align}
    & \Sigma^{(2,2)}_{\hbb,E} = \left( \frac{11}{2}-\pi ^2+\log ^2(2) \right)C_F^2 + \left(\frac{\pi^2}{6}-\frac{35}{36}\right)C_F C_A + \frac{1}{9} C_F n_f T_R \, , \nonumber \\ 
    & \Sigma^{(2,1)}_{\hbb,E} = (-11.4892(1))C_F^2 + (-1.9558(1))C_FC_A + (1.28116(4))C_F n_f T_R \, , \nonumber \\
    & \Sigma^{(2,0)}_{\hbb,E} = (1.483(3))C_F^2 + (8.248(1))C_F C_A + (-2.5621(2))C_F n_f T_R\, ,
\end{align}
while in the WTA scheme we have
\begin{align}
    & \Sigma^{(2,2)}_{\hbb,{\rm WTA}} = C_F^2\left(\frac{11}{2}-\frac{5 \pi ^2}{6}  \right) + \left( \frac{\pi^2}{6} - \frac{35}{36} \right)C_F C_A + \frac{1}{9} C_F n_f T_R \, , \nonumber \\
    & \Sigma^{(2,1)}_{\hbb,{\rm WTA}} = (-6.1943(1))C_F^2 + (-2.84899(4))C_F C_A + (1.38354(4))C_F n_f T_R \, , \nonumber \\
    & \Sigma^{(2,0)}_{\hbb,{\rm WTA}} = (12.643(1))C_F^2 + (5.2378(7)) C_F C_A + (-2.2556(3))C_F n_f T_R \, .
\end{align}

\section{Numerical results}
\label{sec:results}

To complete the evaluation of $d\sigma^{\rm (N)NLO}$ in Eq.~(\ref{eq:xs}), the cross section above the cut, $d\sigma^{\rm (N)NLO}_{\qf>\qfcut}$ is needed. Its computation is performed employing the dipole subtraction formalism~\cite{Catani:1996jh,Catani:1996vz}, within an extension of the {\sc Matrix} framework~\cite{Grazzini:2017mhc} to deal with $e^+e^-$ collisions. The required QCD amplitudes are computed with {\sc OpenLoops}~\cite{Cascioli:2011va,Buccioni:2017yxi,Buccioni:2019sur}. To perform the numerical validation of our results separated into colour channels, we also use the analytic amplitudes provided in Ref.~\cite{Ellis:1980wv} and Ref.~\cite{DelDuca:2015zqa} for \mbox{$\twoj$} and \mbox{$\hbb$}, respectively.

Our calculation is fully differential, allowing the application of arbitrary IR-safe cuts on the final-state partons. However, fully differential NNLO results for \mbox{$\twoj$}~\cite{Anastasiou:2004qd,Gehrmann-DeRidder:2004ttg} and \mbox{$\hbb$}~\cite{Anastasiou:2011qx,DelDuca:2015zqa} have been available for a long time, and the numerical study presented here is intended solely to validate our computational framework and to test the quality of the numerical results.
Accordingly, we limit ourselves to presenting the inclusive NNLO corrections as a function of the slicing parameter \mbox{$\rcut=\qfcut/\sqrt{Q^2}$}, for \mbox{$\twoj$} and \mbox{$\hbb$}, and compare the outcome for each colour channel with the exact analytic results~\cite{Gorishnii:1990zu,Gorishnii:1991zr,Baikov:2005rw}. Two variants of the slicing variable $y_{23}$ are used, employing the $E$-scheme and the WTA scheme, respectively. The results are further compared with those obtained from an independent implementation of thrust~\cite{Farhi:1977sg} $T=1-\tau$, which may be viewed as a version of 2-jettiness for these simple processes.
In this case, the coefficients $\Sigma^{(i,j)}$ are constructed by using the results from Refs.~\cite{Monni:2011gb,Gao:2019mlt} (their explicit expressions are reported in Appendix~\ref{app:thrust}).

When comparing results obtained with different slicing variables, the question of how the comparison should be performed naturally arises.
In the literature, such comparisons are often presented by using $\tau^{1/\sqrt{2}}$, and we use this convention here as well, by showing the quality of the convergence as \mbox{$\rcut\to 0$} for the various variables, obtained using the {\it same} computing resources.

\begin{figure}[t]
\begin{center}
\begin{tabular}{cc}
\includegraphics[width=0.49\textwidth]{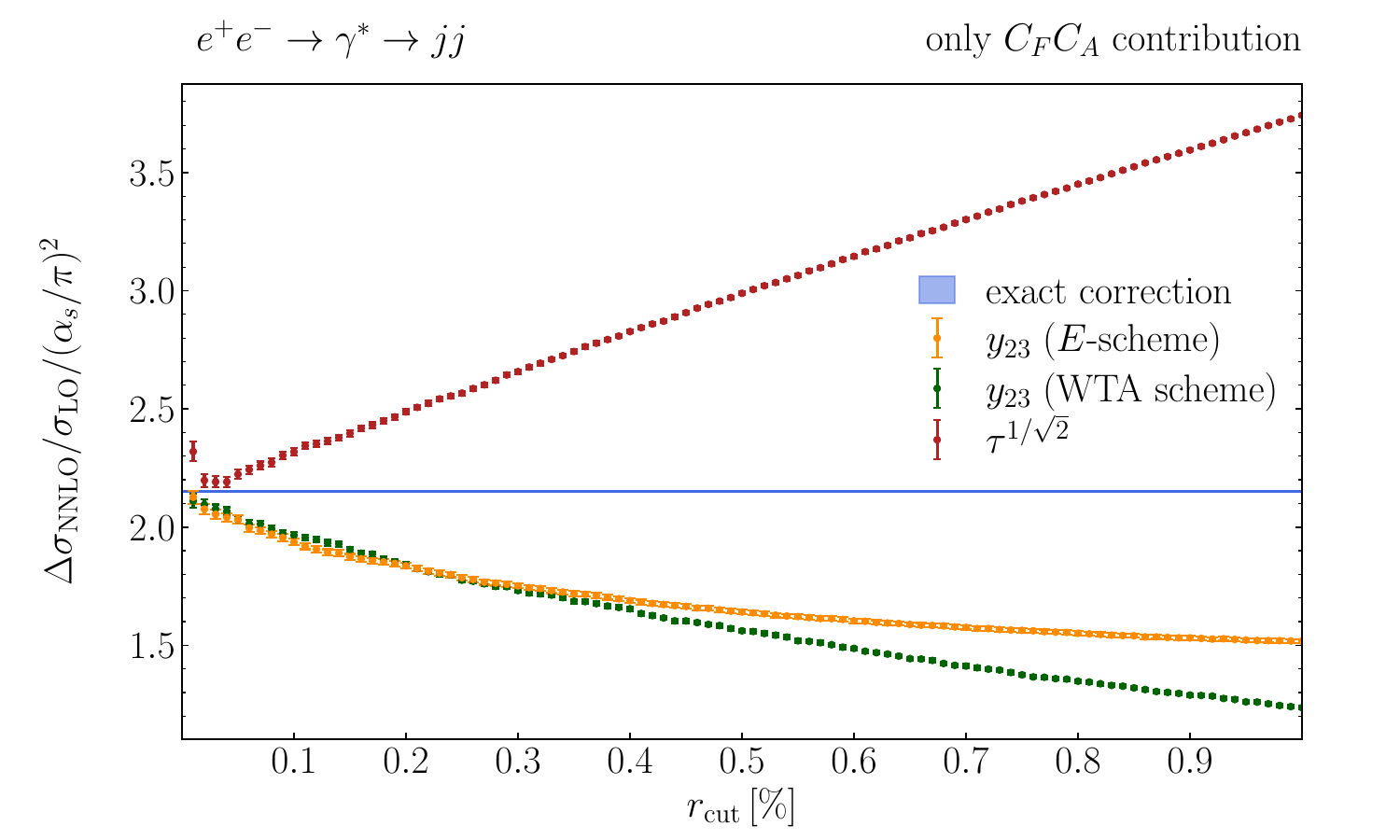} &
\includegraphics[width=0.49\textwidth]{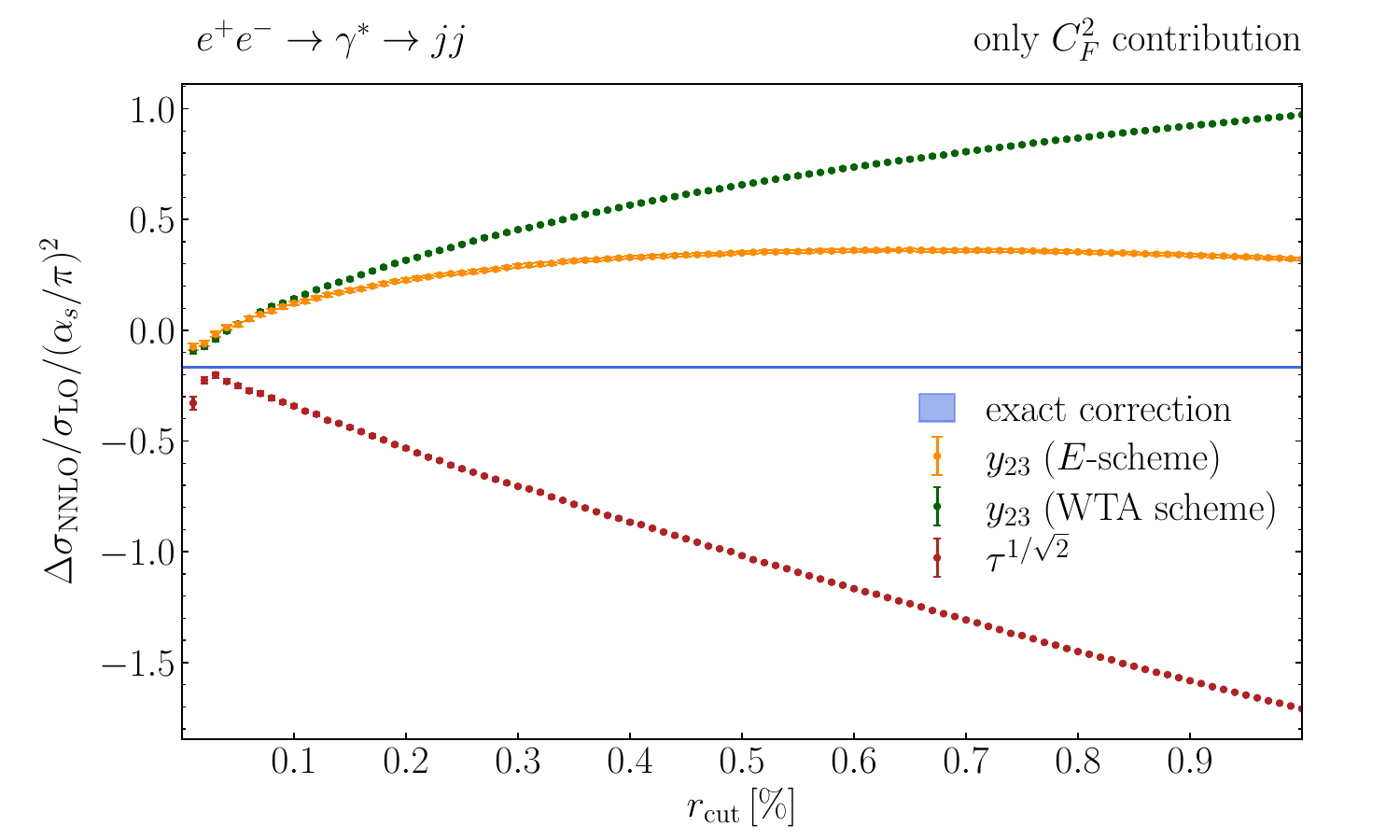}\\
\includegraphics[width=0.49\textwidth]{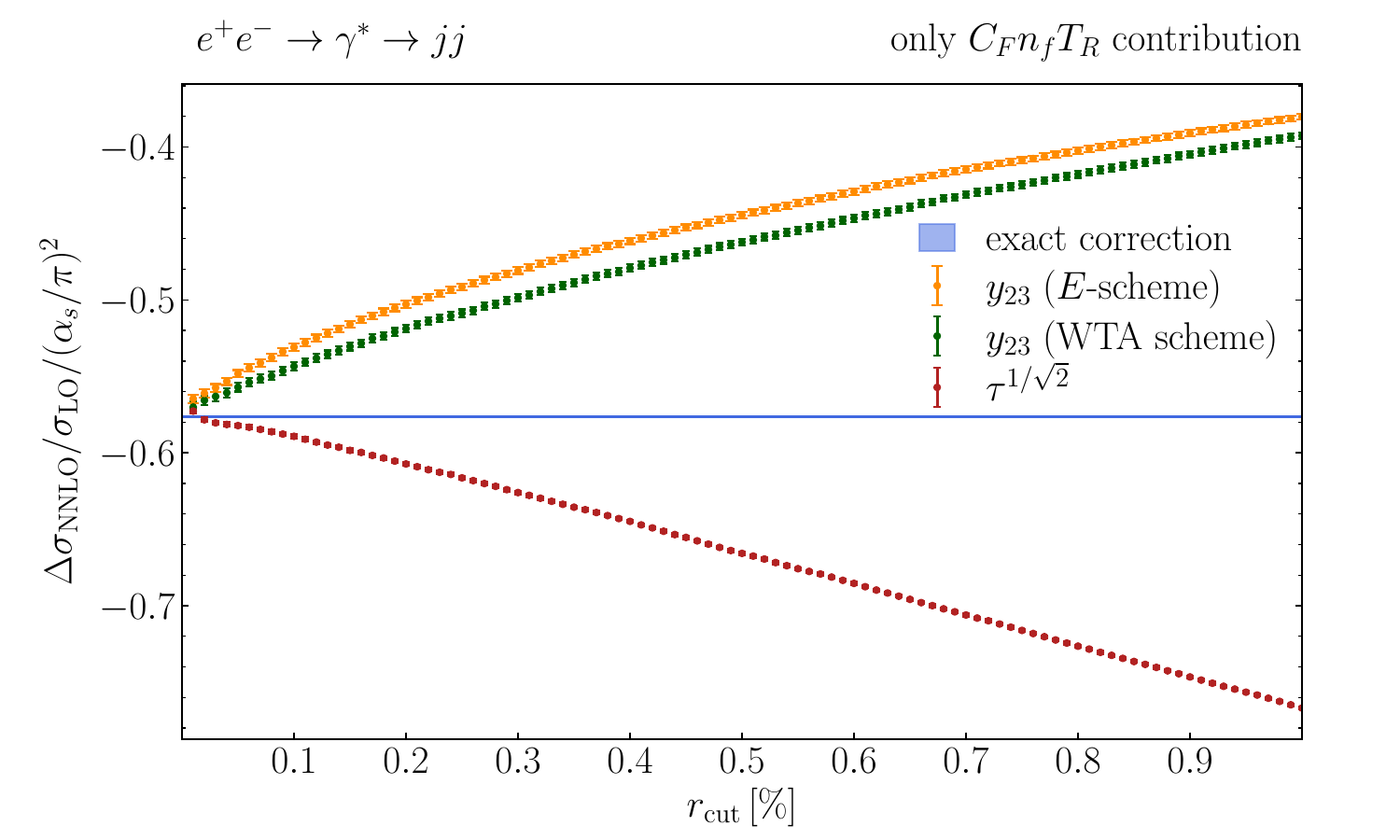} &
\includegraphics[width=0.49\textwidth]{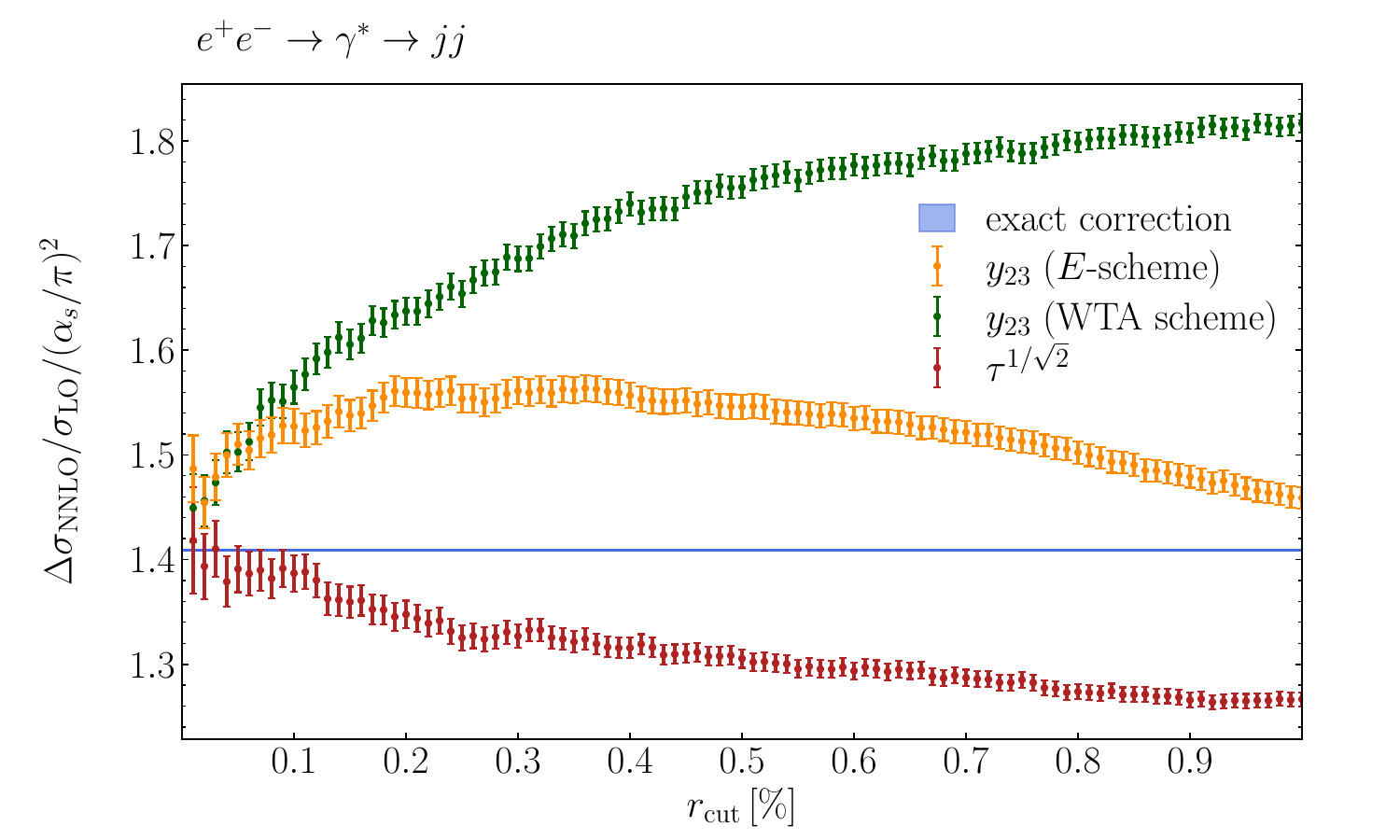}
\end{tabular}
\end{center}
\caption[]{\label{fig:Ajj}{NNLO correction in the various colour channels computed with different variables as a function of the slicing parameter. The exact result is shown in blue.}}
\end{figure}

\begin{figure}[t]
\begin{center}
\begin{tabular}{cc}
\includegraphics[width=0.49\textwidth]{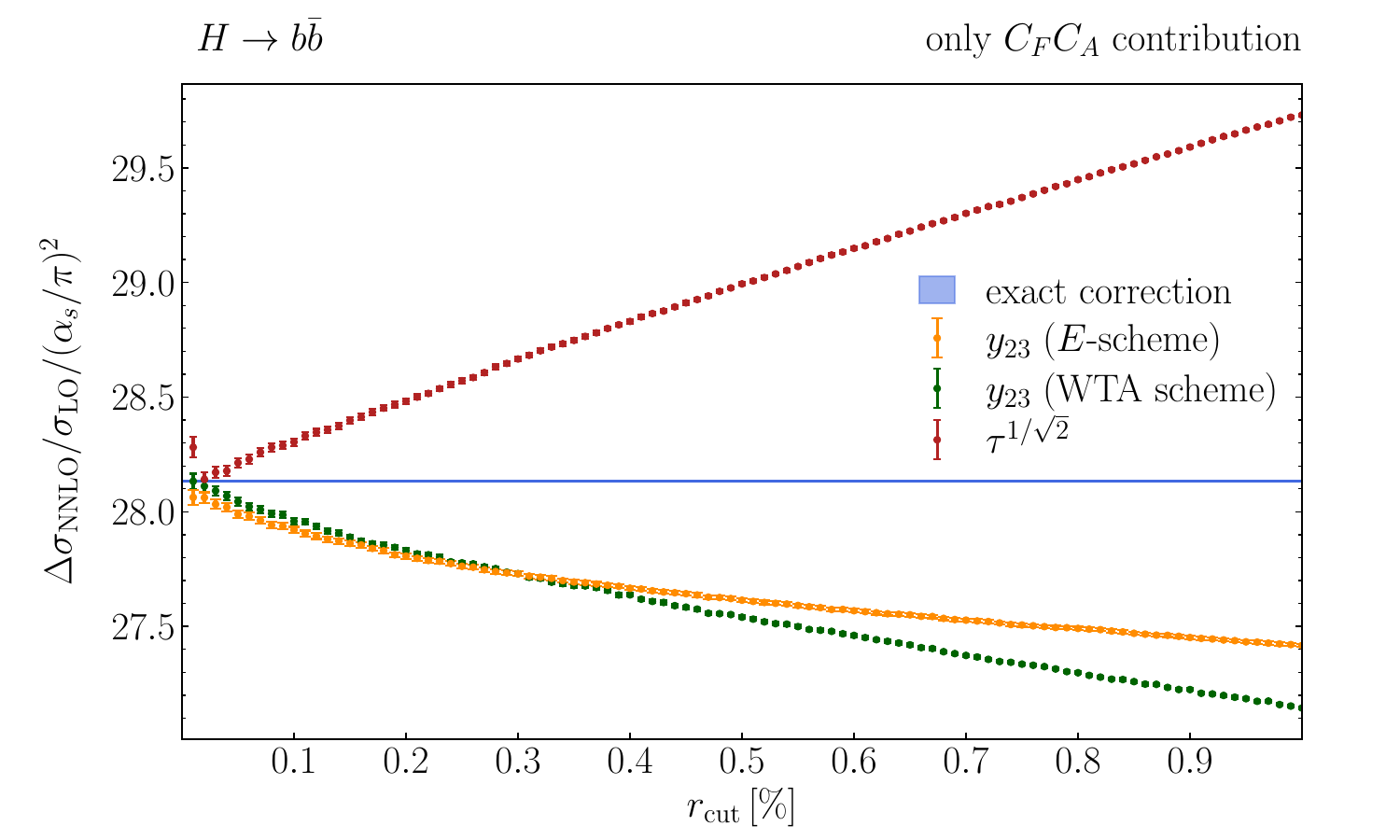} &
\includegraphics[width=0.49\textwidth]{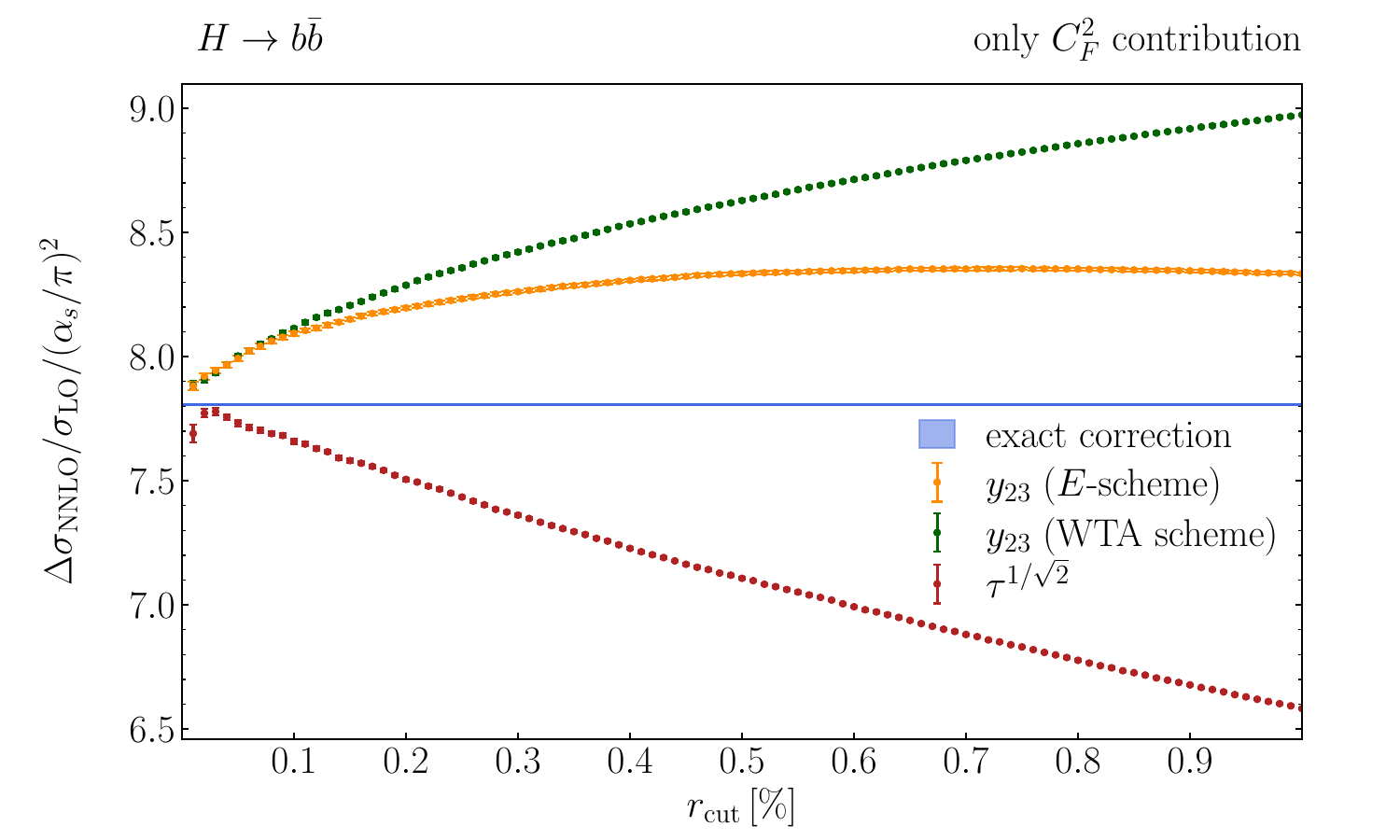}\\
\includegraphics[width=0.49\textwidth]{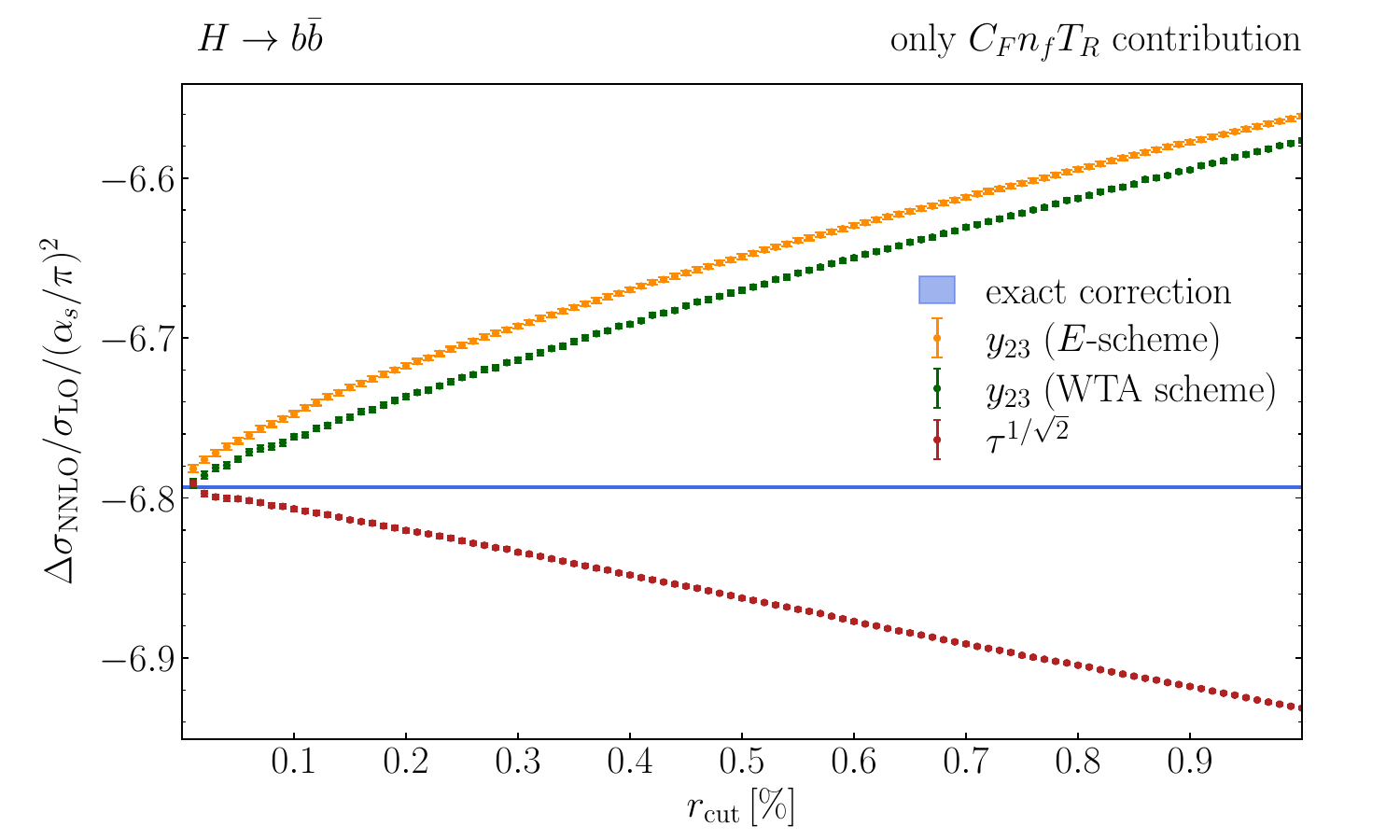} &
\includegraphics[width=0.49\textwidth]{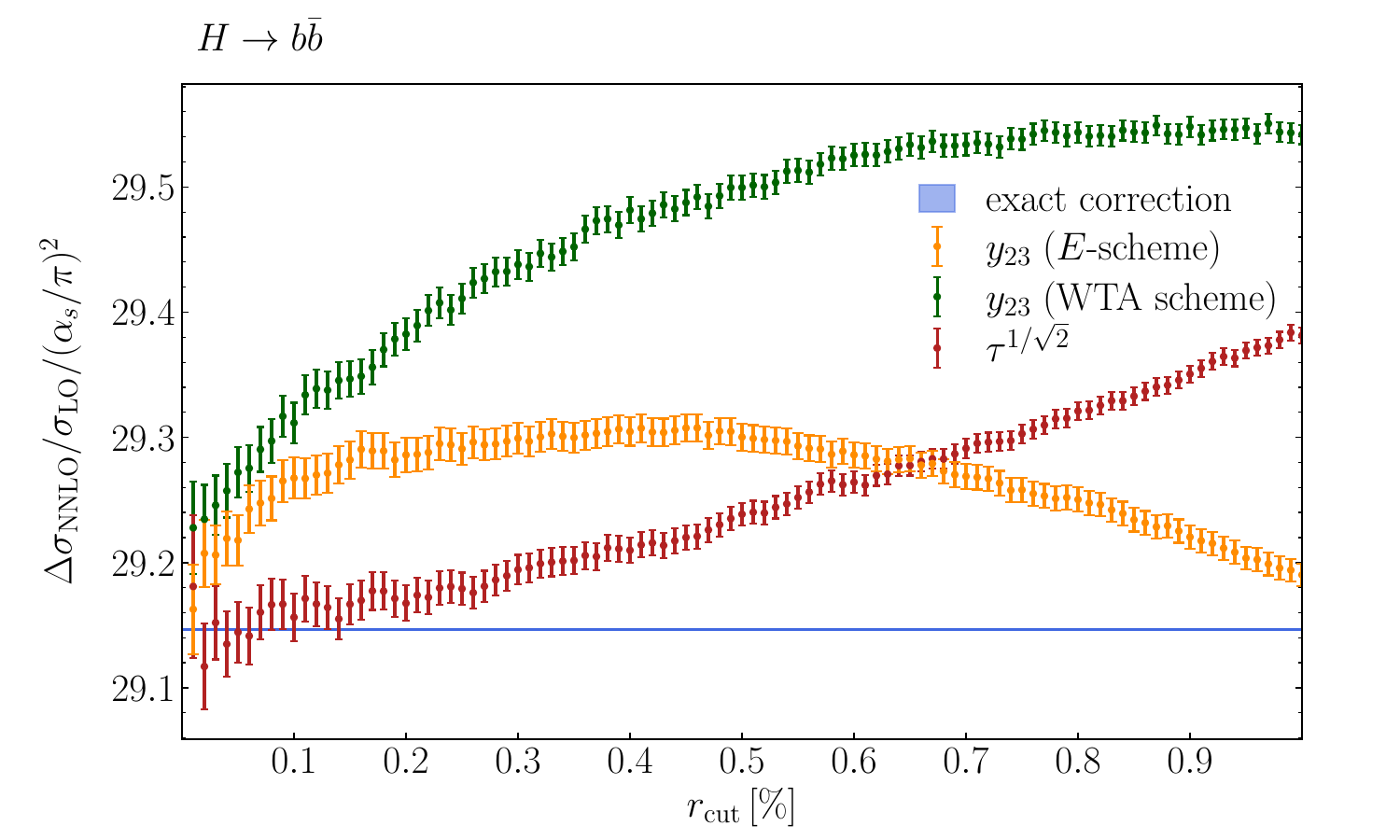}
\end{tabular}
\end{center}
\caption[]{\label{fig:Hbb}{As in Fig.~\ref{fig:Ajj} but for $\hbb$.}}
\end{figure}

The NNLO results are shown in Figs.~\ref{fig:Ajj} and \ref{fig:Hbb}. These results correspond to about $10^5$ CPU hours for each colour structure and each subtraction scheme. Overall we observe a nice convergence of the computed NNLO correction to the exact result for each colour channel.
When considering the full NNLO correction to \mbox{$\twoj$} (bottom-right panel in Fig.~\ref{fig:Ajj}), large cancellations occur among the three contributions, and the precision of the complete correction is only about $2\%$. In the case of \mbox{$\hbb$} we observe a much better control of the complete correction, at the per mille level.
The computational cost mentioned above may seem quite large for such relatively simple processes. Note, however, that the precision achieved at the lowest calculated value \mbox{$\rcut=0.01\%$} is better than $0.05$\textperthousand{} of the NNLO cross section. 
The purpose of this study, and in particular of the splitting into colour structures, is to numerically validate all ingredients of our calculation. The overall precision of the complete result is widely dominated by the error on the $C_FC_A$ channel, the $C_F^2$ error is smaller by a factor of two and that from $C_Fn_f T_R$ by an order of magnitude, with comparable runtimes for each colour channel.
Individually, this level of precision corresponds to control over the above-cut contribution at the level of about \mbox{$4\times10^{-6}$} for the $C_FC_A$ channel, and better than $10^{-6}$ for the other channels.

Comparing the results obtained with the two variants of $y_{23}$, we note that the behaviour is very similar in the $C_FC_A$ and $C_Fn_f T_R$ channels. In the $C_F^2$ contribution, the $\rcut$ dependence in the $E$-scheme appears to be somewhat flatter. The latter comment applies to both \mbox{$\twoj$} and \mbox{$\hbb$}. Comparing the $y_{23}$ results with those obtained with 2-jettiness, we note that the $\rcut$ dependence is opposite in all the colour channels. As for the quality of the convergence to the exact result, we do not see significant differences, and the uncertainties in a \mbox{$\rcut\to 0$} extrapolation would be quite similar.

\section{Summary}
\label{sec:summa}
In this paper we have discussed a new non-local subtraction scheme suitable for NNLO computations with final-state jets. The formalism is based on the use of transverse-momentum--like variables as slicing variables to handle and cancel IR singularities. As a first application, we explicitly considered dijet production in $e^+ e^-$ collisions and \mbox{$H\rightarrow b \bar b$} decays in NNLO QCD. We have discussed the perturbative ingredients required for the computation. The quark jet function has been presented recently~\cite{Buonocore:2025ysd}. The soft function has been computed using two variants of $y_{23}$ in which jets are reconstructed with the $E$-scheme and the WTA scheme, respectively. Using the $E$-scheme, a factorisation-breaking term emerges at ${\cal O}(\as^2)$, and we have explicitly computed it.

Our calculation has been implemented in the {\sc Matrix} framework, which, for the first time, has been extended to deal with this class of processes.  We have presented numerical results for the NNLO corrections using the two variants of $y_{23}$ and compared them with an independent implementation of the thrust variable, which, for these simple processes, may be viewed as a version of 2-jettiness.
We have validated our implementation by breaking the results into the three colour channels, comparing each result with the corresponding analytical coefficient for the inclusive NNLO cross section.
For both processes, the achieved precision at the level of the NNLO cross section is excellent, better than $0.1$\textperthousand{}. This corresponds to a precision of the NNLO coefficient at the per mille level for \mbox{$\hbb$}, and of about $2\%$ for $\twoj$, due to large accidental cancellations between the various colour channels.

Comparing the performance of the three slicing variables, we do not find significant differences.
It will be interesting to see whether this conclusion is modified for higher jet multiplicities.
The extension of our subtraction framework to more complicated processes requires additional perturbative ingredients that are currently unavailable. For example, to compute NNLO corrections to $e^+e^-\to 3\,{\rm jets}$, the gluon jet function is needed.

We look forward to extending these results to more complicated processes.

\noindent {\bf Acknowledgements}. We would like to thank Prasanna Dhani for discussions and comments on the manuscript.
This work is supported in part by the Swiss National Science Foundation (SNSF)
under contracts 200020$\_$219367 and 200021$\_$219377. The work of L.B. is funded by the European
Union (ERC, grant agreement No. 101044599, JANUS). Views and opinions expressed
are however those of the authors only and do not necessarily reflect those of
the European Union or the European Research Council Executive Agency. Neither
the European Union nor the granting authority can be held responsible for them.

\appendix

\section{Hard function}
\label{app:hard}

The coefficients controlling the expansion of the hard function in Eq.~\eqref{eq:exp} up to ${\cal O}(\as^2)$ can be
obtained from the one- and two-loop amplitudes \mbox{$\gamma^*\to q\bar{q}$}~\cite{Matsuura:1988sm} and \mbox{$\hbb$}~\cite{Ravindran:2006cg}.
Using the notation 
\begin{equation}
H^{(1)} = \frac{1}{4}C_F h_F, \quad\quad H^{(2)} = \frac{1}{16}\left(C_F^2 h_{2F} + C_F C_A h_A + C_F n_f T_R h_f\right)\, ,
\end{equation}
the coefficients $h_F$, $h_{2F}$, $h_A$ and $h_f$ for $e^+e^-\to q\bar{q}$ read
\begin{align}
    h_F &= -\frac{4}{\eps^2} -\frac{6}{\eps} -16 + \frac{7\pi^2}{3} + \eps \left( -32+\frac{7\pi^2}{2} + \frac{28\zeta_3}{3} \right)\! + \eps^2 \left( -64 + \frac{28\pi^2}{3} + 14\zeta_3 - \frac{73\pi^4}{360} \right)\! \, , \nonumber \\
    h_{2F} & = \frac{8}{\eps^4} + \frac{24}{\eps^3} + \frac{1}{\eps^2} \!\left( 82-\frac{28\pi^2}{3} \right)\! + \frac{1}{\eps}\! \left( \frac{445}{2} -26\pi^2 -\frac{184\zeta_3}{3} \right)\! + \frac{2303}{4} - 86\pi^2 -172\zeta_3 +\frac{137\pi^4}{45} \, , \nonumber \\
    h_A & = -\frac{11}{3\eps^3} + \frac{1}{\eps^2} \!\left(-\frac{166}{9} + \frac{\pi^2}{3} \right)\! + \frac{1}{\eps}\!\left(-\frac{4129}{54} + \frac{121 \pi^2}{18} + 26 \zeta_3\right)\! - \frac{89173}{324} + \frac{877 \pi^2}{27} + \frac{934\zeta_3}{9} - \frac{8\pi^4}{45} \, , \nonumber \\
    h_f & = \frac{4}{3\eps^3} + \frac{56}{9\eps^2} + \frac{1}{\eps}\left( \frac{706}{27} - \frac{22\pi^2}{9} \right) + \frac{7541}{81} - \frac{308 \pi^2}{27} - \frac{104\zeta _3}{9}\, .
\end{align}
For the decay \mbox{$\hbb$} they read
\begin{align}
    h_F &=  -\frac{4}{\eps^2} -\frac{6}{\eps} -12 L_H + \frac{7\pi^2}{3} -4+\eps\left( -12L_H^2 + \frac{28\zeta_3}{3} - 8\right) + \eps ^2 \left( -8L_H^3 -16 + \frac{7\pi^2}{3} - \frac{73\pi^4}{360} \right) \, , \nonumber \\
    h_{2F} &=  \frac{8}{\eps^4} + \frac{24}{\eps^3} + \frac{1}{\eps^2}\left( 48 L_H +34 -\frac{28\pi^2}{3}\right) + \frac{1}{\eps} \left( 48L_H^2 + 72 L_H -\frac{184\zeta_3}{3} + \frac{109}{2} - 12\pi^2 \right)  + \nonumber \\
    & + 32L_H^3 + 144 L_H^2 + L_H (42-28\pi^2) - 116 \zeta_3 + 128 -\frac{40\pi^2}{3} + \frac{137 \pi^4}{45} \, , \nonumber \\
    h_A &= -\frac{11}{3\eps^3} + \frac{1}{\eps^2} \left( \frac{\pi^2}{3} - \frac{166}{9} \right) + \frac{1}{\eps} \left( -44L_H + 26 \zeta_3 - \frac{1753}{54} + \frac{121\pi^2}{18} \right) - 88 L_H^2 - \frac{194}{3}L_H \nonumber \\
    & +\frac{610\zeta_3}{9} - \frac{3310}{81} + \frac{701\pi^2}{54} - \frac{8\pi^4}{45} \, , \nonumber \\
    h_f &= \frac{4}{3\eps^3} + \frac{56}{9\eps^2} + \frac{1}{\eps} \left( 16L_H + \frac{274}{27} -\frac{22\pi^2}{9} \right) + 32 L_H^2 + \frac{40}{3} L_H - \frac{104\zeta_3}{9} + \frac{1664}{81} - \frac{110\pi^2}{27}\, ,
\end{align}
where we defined \mbox{$L_H=\ln(m_H/\mu)$}, $m_H$ being the mass of the Higgs boson.

\section{Relevant soft and collinear limits}
\label{app:limits}

The calculation of the subtracted soft functions of Eqs.~\eqref{eq:softqq}, \eqref{eq: softggnab} and~\eqref{eq:dsoftnab} 
requires the expansion of the observable $\qf$ in various kinematical limits. In particular, we have to evaluate expressions of the type $\qf_A$, $\qf_{A,B}$, $\qf_{A,B,C}$, where $\qf_{A,B}$ means that the observable is expanded first 
in the limit $A$ and then in the limit $B$, and similarly for $\qf_{A,B,C}$. When the observable is evaluated in mixed 
soft and collinear limits, we will use the notation $\qf_{AB}(\{k\},\{l\})$ (without comma in the $AB$ subscript), which means that $\qf$ is expanded 
considering the set of momenta $\{k\}$ in the limit $A$, and the set of momenta $\{l\}$ in the limit $B$. For example, the expression $\qf_{C_iS}(\{k_1,k_2\},\{k_3\})$ means that we have to expand the observable in the limit in which $k_1$ and $k_2$ are collinear to the hard momentum 
$p_i$, and $k_3$ is soft. The function $\qf_{C_iS,S}(k_2,k_3)$ is then obtained from $\qf_{C_iS}(\{k_1,k_2\},\{k_3\})$ by taking $k_2$ and $k_3$ soft. A detailed explanation of how to perform these expansions is provided in Appendices~D and E of Ref.~\cite{Haag:2025ywj}.

Before giving the expressions of $\qf$ in the kinematical limits, we discuss some subtleties appearing when expanding the distance 
$d_{ij}$ of Eq.~\eqref{eq:dijdef}: 
\begin{equation}
  d_{ij}^2 = \frac{E_i^2 E_j^2}{(E_i + E_j)^2} 2 (1-\cos \vartheta _{ij}) \, .
\end{equation}

Let us consider for example the approximation $\qf_{C_1S}(\{k_1,k_2\},\{k_3\})$. To decide if $k_3$ is clustered with $k_1$ or $k_2$, we have to evaluate the Heaviside function \mbox{$\theta (d_{13}-d_{23})$}. At leading power, the soft particle is only sensitive to the direction of the hard momentum $p_i$, so we have 
\begin{equation}
  d^2_{13} \simeq d^2_{23} = 2 E_3^2 (1-\cos \vartheta ) + \dots\,,
\end{equation}
where the dots represent subleading terms, and the angle $\vartheta$ refers to the angle between $k_3$ and the hard momentum $p_1$. Thus, to evaluate the Heaviside function to leading power, we have to expand the difference \mbox{$d_{13}-d_{23}$} to the first subleading power. We denote the first subleading-power expansion of $d_{ij}$ by $\tilde{d}_{ij}$. This expansion gives
\begin{align}
\label{eq: theta tilde}
   \theta (\tilde{d}_{13} - \tilde{d}_{23}) = \theta \bigg( (z_1-z_2)(e^{-2y_3} + 1) + \frac{k_{2,\perp}}{k_{3,\perp}} \cos \varphi _{23} \bigg) \, , 
\end{align}
where \mbox{$y_3 = \frac{1}{2}\log\left( \frac{1+\cos\vartheta}{1-\cos\vartheta} \right)$} is the rapidity of $k_3$, \mbox{$k_{\perp,2}=k_{\perp,1}$} the transverse momentum of $k_2$, and $\varphi_{23}$ the angle between the transverse momentum $k_{\perp,3}$ and $k_{\perp,2}$, all with respect to the hard quark momenta $p_1$ and $p_2$.

Something similar happens in the function $\qf_{C_i,C_iS}(\{k_1,k_2\},\{k_3\})$, where the variable is 
first expanded in the triple collinear limit \mbox{$k_1\parallel k_2 \parallel k_3$} along the direction of $p_i$, 
and then the momentum $k_3$ becomes soft, while $k_1$ and $k_2$ are kept collinear. In the first expansion, 
the distances $d_{jk}$ are evaluated in the collinear limit. We denote the leading-power approximation of the distance $d_{jk}$ in the collinear limit along $p_i$ as $d_{C_i,jk}$. We find
\begin{equation}
  d_{C_i,jk}^2 = \frac{z_j^2 z_k^2}{(z_j + z_k)^2} \bigg(\frac{k_{j,\perp}^2}{z_j^2} + \frac{k_{k,\perp}^2}{z_k^2} - 2\frac{k_{j,\perp} k_{k,\perp}}{z_j z_k} \cos \varphi _{jk} \bigg) \, ,
\end{equation}
where $z_j$, $z_k$ are the longitudinal momentum fractions of the collinear particles with respect to the direction $p_i$, and $\varphi_{jk}$ is the angle between the transverse momenta $k_{\perp,j}$ and $k_{\perp,k}$.
When we take the soft limit \mbox{$k_3\to 0$} while keeping $k_1$ and $k_2$ collinear, we have again to perform a subleading-power expansion to determine if $k_3$ is clustered with $k_1$ or $k_2$. We write the first subleading-power expansion of $d_{C_i,jk}$ as $\tilde{d}_{C_i,jk}$. The Heaviside function \mbox{$\theta (\tilde{d}_{C_i,13} - \tilde{d}_{C_i,23})$} yields
\begin{align}
\label{eq: thetac tilde}
  \theta (\tilde{d}_{C_i,13} -  \tilde{d}_{C_i,23})=\lim_{y_3\to \infty} \theta (\tilde{d}_{13} - \tilde{d}_{23}) = \theta \bigg(z_1-z_2 + \frac{k_{2,\perp}}{k_{3,\perp}} \cos \varphi _{23} \bigg) \, .
\end{align}
When computing the limits $\qf_{C_iS,S}$ and $\qf_{C_i,C_iS,S}$, the soft limit 
\mbox{$z_2 \to 0$} of Eqs.~\eqref{eq: theta tilde} and~\eqref{eq: thetac tilde} has to be taken. 
We will denote this additional soft limit of the subleading-power expansion of the distances $\tilde{d}_{ij}$ and $\tilde{d}_{C_i,jk}$ as 
$\tilde{\tilde{d}}_{ij}$ and $\tilde{\tilde{d}}_{C_i,jk}$, respectively. The soft limits of the Heaviside functions 
read
\begin{align}
   \theta (\tilde{\tilde{d}}_{13} - \tilde{\tilde{d}}_{23}) &= \lim_{z_2\to0}\theta (\tilde{d}_{13} - \tilde{d}_{23}) =\theta \bigg(1+e^{-2y_3} + \frac{k_{2,\perp}}{k_{3,\perp}} \cos \varphi _{23} \bigg) \, , \nonumber \\
  \theta (\tilde{\tilde{d}}_{C_i,13} - \tilde{\tilde{d}}_{C_i,23}) & =\lim_{z_2\to 0}\theta (\tilde{d}_{C_i,13} - \tilde{d}_{C_i,23}) =\theta \bigg(1 + \frac{k_{2,\perp}}{k_{3,\perp}} \cos \varphi _{23} \bigg) \, .
\end{align}

We now report the expressions of the observable in the various kinematical limits:
\begin{align}
     \qf_{C_iS}(\{ k_1,  k_2\}, \{ k_3\})&= \max\left(\qf_{C_i}( k_1,  k_2), \qf_S(k_3)\right)\notag\\
     &\hspace{-3.5cm} +\alpha \, \theta\!\left(\pm\cos\vartheta-\max\left( 1-\frac{  k_{1, \perp}^2}{2E_3^2}, 0 \right) \right)\left[ \theta\!\left( \tilde{d}_{13}-\tilde{d}_{23} \right)\left( d_{C_i, 1(2+3)}- \qf_{C_i}( k_1,  k_2) \right)+ \left( 1\leftrightarrow 2 \right)\right] \, , \nonumber \\
     \qf_{C_iC_j, C_iS}(\{k_1, k_2\}, \{k_3\})&=\max\left( \qf_{C_i}(k_1, k_2),   k_{3,\perp}  \right)\, , \notag\\
     \qf_{C_i, C_iS}(\{k_1, k_2\}, \{k_3\})&=\max \left(\qf_{C_i}(k_1, k_2),   k_{3, \perp}   \right)\notag\\
    &+\alpha \,\theta\!\left(   k_{1, \perp}  -  k_{3, \perp} \right)\left[ \theta(\tilde{d}_{C_i, 13}-\tilde{d}_{C_i, 23} )\left( d_{C_i, 1(2+3)} -  \qf_{C_i}( k_1,  k_2)  \right)+\left( 1\leftrightarrow 2 \right)\right]\, , \notag\\
    \qf_S(k_1, k_2)&=\max(\qf_{S}(k_1), \qf_{S}(k_2)) \notag\\
    &\hspace{-1.5cm} + \theta\!\left( \min\left(  \qf_S(k_1), \qf_S(k_2)\right)-d_{12}\right)\left[\qf_S(k_{12})-\max\left( \qf_S(k_1), \qf_S(k_2) \right) \right]\notag\, , \\
    \qf_{C_i, S}(k_2, k_3)&=\theta(\min(  k_{2, \perp} ,   k_{3,\perp} )-d_{C_i, 23})  k^{C_i}_{23,\perp}  \notag\\
        &+\theta(d_{C_i, 23}-\min(  k_{2, \perp} ,   k_{3,\perp} )) \max(  k_{2, \perp} ,   k_{3,\perp} )\notag\, , \\
    \qf_{C_iS, S}(k_2, k_3)&=\max\left(k_{2,\perp}, \qf_S(k_3)\right)\notag \\
    &\hspace{-1.5cm}+\alpha\,\theta\!\left(\pm\cos\vartheta-\max\left( 1-\frac{  k_{2, \perp} ^2}{2E_3^2}, 0 \right) \right) \theta\!\left(\tilde{ \tilde{d}}_{13}-\tilde{\tilde{d}}_{23} \right)\left(   k^{C_i}_{23,\perp}  -  k_{2, \perp}  \right)\notag\, , \\
    \qf_{C_iC_j, C_iS, S}(k_2, k_3)&=\qf_{C_iC_j, SC_j, S}(k_2, k_3)=\max\left(   k_{2, \perp} ,   k_{3 ,\perp}  \right)\notag \, , \\
    \qf_{C_i, C_iS, S}(k_2, k_3)&=\qf_{C_i, SC_i, S}(k_3, k_2)=\max\left(   k_{2, \perp} ,   k_{3 ,\perp}  \right)+\notag\\
    &+\alpha\,\theta\!\left(   k_{2, \perp}  -  k_{3, \perp} \right)\theta(\tilde{\tilde{d}}_{C_i, 13}-\tilde{\tilde{d}}_{C_i, 23} )\left(   k^{C_i}_{23,\perp}  -  k_{2, \perp}  \right)\, .
  \end{align}
To obtain the $E$-scheme result, set \mbox{$\alpha=1$}; for the WTA scheme set \mbox{$\alpha=0$}. 
The $\pm$ sign in the theta functions is $+$ for $C_1$ and $-$ for $C_2$. The angle $\vartheta$ refers to the angle between $k_3$ and $p_1$. Note that the momentum fractions are defined differently depending on the collinear direction: 
 \begin{equation}
  z_j=\begin{cases}
  \frac{k_j\cdot p_2}{p_1\cdot p_2} & \text{for }C_i=C_1\,,\\
  \frac{k_j\cdot p_1}{p_1\cdot p_2} & \text{for }C_i=C_2\, .
  \end{cases}
 \end{equation}
 The four momentum $k^{C_i}_{23,\perp} $ refers to the transverse part of the momentum obtained by clustering $k_2$ and $k_3$ in the $C_i$ collinear limit. It is given by
 \begin{equation}
 k^{C_i,\mu}_{23,\perp}= \begin{cases} k_{2,\perp}^\mu+k_{3,\perp}^\mu, & \text{ for $E$-scheme,}\\
    \frac{z_2 + z_3}{z_2} k_{2, \perp}^\mu & \text{ if } z_2 > z_3 \text{ for WTA scheme,}\\
    \frac{z_2 + z_3}{z_3} k_{3 ,\perp}^\mu & \text{ if } z_3 > z_2 \text{ for WTA scheme.}
  \end{cases}
 \end{equation}
 For the $E$-scheme one further needs the distances $d_{C_i, 1(2+3)}$ and $d_{C_i, 2(1+3)}$, where a soft parton $k_3$ clusters with either one of the two collinear partons $k_1$ or $k_2$. They read\footnote{In the WTA scheme we obtain $d_{C_i, 1(2+3)}=d_{C_i, 2(1+3)}=  \qf_{C_i}( k_1,  k_2)$, leading to $d_{C_i, 1(2+3)}-d_{C_i, 2(1+3)}=0$, which is why the formulas look simpler in that case.}
\begin{equation}
   \begin{aligned}
    d_{C_i, 1(2+3)}&=\sqrt{   k_{1,\perp} ^2+2 \cos\varphi_{23}    k_{\perp ,1}    k_{3 ,\perp}  z_1+  k_{3 ,\perp} ^2 z_1^2}\,,\\
      d_{C_i, 2(1+3)}&=\sqrt{   k_{1,\perp} ^2-2 \cos\varphi_{23}    k_{\perp ,1}    k_{3 ,\perp}  z_2+  k_{3 ,\perp} ^2 z_2^2}\, .
   \end{aligned}
\end{equation}

\section{Results for $d\sigma_{\qf<\qfcut}$ for thrust}
\label{app:thrust}

For completeness, in this appendix we report the expression for $d\sigma _{\qf < \qfcut}$ for thrust.

At NLO the unresolved cross section for dijet production in $e^+e^-$ collisions is given by 
\begin{equation}
\Sigma^{(1)}_{\twoj}=C_F \left(-L^2+\frac{3}{2} L+\frac{\pi ^2}{6}-\frac{1}{2}\right),
\end{equation}
whereas for the decay \mbox{$\hbb$} the result is
\begin{equation}
    \Sigma ^{(1)}_\hbb =C_F \left(-L^2+\frac{3}{2} L+\frac{\pi ^2}{6}+\frac{5}{2}\right),
\end{equation}
with \mbox{$L=\ln(1/\tau_{\rm cut})$}.

At NNLO we have
\begin{align}
\Sigma^{(2)}_{\twoj}&=
 C_F^2\left[\frac{1}{2} L^4-\frac{3}{2} L^3+\left(\frac{13}{8}-\frac{\pi ^2}{2}\right) L^2 +\left(-\zeta_3-\frac{9}{16}+\frac{\pi ^2}{2}\right) L \right. \nonumber \\
   & \left. \qquad\quad -\frac{3 \zeta_3}{2}+\frac{5 \pi ^4}{144}-\frac{3 \pi
   ^2}{32}+\frac{1}{4}\right] \nonumber \\
    &
    +	C_A C_F \left[-\frac{11}{12} L^3+\left(\frac{\pi
   ^2}{12}-\frac{169}{144}\right)L^2 +\left(\frac{57}{16}-\frac{3 \zeta_3}{2}\right)L \right. \nonumber \\
   & \left. \qquad \qquad +\frac{71 \zeta_3}{12}+\frac{11 \pi ^4}{1440}-\frac{53 \pi
   ^2}{432}-\frac{61}{48}\right]\nonumber \\
   & + C_F n_f T_R \left(\frac{1}{3} L^3+\frac{11}{36} L^2-\frac{5}{4} L-\frac{4 \zeta_3}{3}+\frac{7
   \pi ^2}{108}+\frac{1}{12}\right)
\end{align}
for \mbox{$\twoj$} and
\begin{align}
\Sigma^{(2)}_{\hbb}&=
 C_F^2\left[\frac{1}{2} L^4-\frac{3}{2} L^3+\left(-\frac{11}{8}-\frac{\pi ^2}{2}\right) L^2  +\left(-\zeta_3+\frac{63}{16}+\frac{\pi ^2}{2}\right) L \right. \nonumber \\
   & \left.\quad\quad\quad -\frac{3 \zeta_3}{2}+\frac{5 \pi ^4}{144}+\frac{13 \pi
   ^2}{96}+\frac{241}{64}\right] \nonumber \\
    &
    +	C_A C_F \left[-\frac{11}{12} L^3+\left(\frac{\pi
   ^2}{12}-\frac{169}{144}\right) L^2+\left(\frac{57}{16}-\frac{3 \zeta_3}{2}\right)L\right. \nonumber \\
    &
 \qquad \qquad \left. + \frac{11 \zeta_3}{3}+\frac{11 \pi ^4}{1440}-\frac{233 \pi
   ^2}{432}+\frac{4537}{576}\right]\nonumber \\
   & + C_F n_f T_R \left(\frac{1}{3} L^3+\frac{11}{36} L^2-\frac{5}{4} L-\frac{4 \zeta_3}{3}+\frac{25
   \pi ^2}{108}-\frac{353}{144}\right)
\end{align}
for \mbox{$\hbb$}.

\bibliography{biblio}

\end{document}